\documentclass[11pt,cits]{JHEP3c}

\usepackage{epsfig}
\usepackage{latexsym}
\usepackage{graphicx}
\usepackage{amsmath,amssymb}
\usepackage{epstopdf}  
\def\unit{\relax{\rm 1\kern-.26em I}}

\def\bea{\begin{eqnarray}}
\def\eea{\end{eqnarray}}
\def\be{\begin{equation}}
\def\ee{\end{equation}}
\def\nn{\nonumber}

\def\Z{{\bf Z}}
\def\al{\alpha^\prime}

\newcommand{\gsim}{\lower.7ex\hbox{$\;\stackrel{\textstyle>}{\sim}\;$}}
\newcommand{\lsim}{\lower.7ex\hbox{$\;\stackrel{\textstyle<}{\sim}\;$}}

\title{Moduli Stabilization in Meta-Stable Heterotic Supergravity  Vacua}
\author{M. Serone and A. Westphal\\
International School for Advanced Studies (SISSA/ISAS)
and INFN, Trieste, Italy\\
E-Mail: \email{serone@sissa.it, westphal@sissa.it}}

\abstract{We revisit the issue of moduli stabilization 
in a class of ${\cal N}=1$ four dimensional supergravity theories which are low energy descriptions
of standard perturbative heterotic string vacua  compactified on Calabi--Yau spaces.
In particular, we show how it is possible to stabilize the universal dilaton and K\"ahler moduli
in a  de Sitter/Minkowski vacuum with low energy supersymmetry breaking 
by means of non--perturbative gauge dynamics, including 
recent results by Intriligator, Seiberg and Shih.
The non--SUSY vacua are meta--stable but sufficiently long--lived.
}

\preprint{SISSA-38/2007/EP\\ July 3, 2007}


\begin{document}

\section{Introduction}

One of the main problems that string theorists have to face when attempting to  construct 
(semi)-realistic string models is the issue of how to fix the Vacuum Expectation Values (VEV's) of 
neutral scalar fields with perturbatively flat potentials, the so called moduli. 
The absence of a potential for the moduli is typically due to supersymmetry (SUSY), which seems to be
a fundamental ingredient (at least at Planckian energies)
 to construct stable string theory vacua.  The problems of how to stabilize moduli and 
 how to break SUSY at low energy in a full--fledged string set--up are among
 the most important unsolved problems in string theory. All the coupling constants of the effective theories describing the low energy
dynamics of string vacua depend on the moduli VEV's, so that it is crucial to address the problem
of how these VEV's are dynamically fixed, in order to establish a connection between string theory
and known particle physics.

In recent years great progress has been achieved in stabilizing the moduli, mainly in 
Type II strings, thanks to the introduction of fluxes for various Ramond--Ramond tensor field strengths
and for the three form torsion field strength $H$ (see \cite{Douglas:2006es} for a recent review and references therein).  We still do not have a tractable string description of such vacua, but at least a low energy description seems possible. In particular, this is sufficiently reliable if the flux back-reaction of the geometry can be neglected, and the compactified space can be taken as unperturbed.\footnote{In fact, the word ``moduli stabilization'' is an artifact of this approximation. Since the flux--induced superpotential is a tree-level effect, strictly speaking, in such compactifications, the would--be moduli are simply not there and the term ``moduli stabilization'' is an abuse of language.} 

The situation in heterotic strings is  more complicated, since the flux for $H$ leads necessarily to non--K\"ahler manifolds \cite{Strominger}, whose properties are still little known (see e.g. \cite{LopesCardoso:2002hd,Becker:2003yv,Brustein:2004xn,deCarlos:2005kh} for some progress in this direction). This is unfortunate because ---  despite the by now
many string vacua built using D--branes in various contexts ---  the old fashioned perturbative
heterotic string vacua on Calabi--Yau manifolds remain one of the most attractive scenarios
for model building. Indeed, this is the ideal context to build SUSY GUT models (see 
\cite{Kobayashi:2004ud,Buchmuller:2005jr,Bouchard:2005ag,Lebedev:2006kn} for recent
constructions), contrary to D--brane model building, where gauge coupling unification is generally lost.

These considerations motivated us to revisit the issue of moduli stabilization in perturbative
(fluxless) heterotic string vacua compactified down to four dimensions on a Calabi--Yau manifold
 \cite{Gross:1985fr}.
In absence of a tree-level induced flux superpotential, one has to rely on non--perturbatively generated superpotentials for the moduli, arising from strong coupling 
gauge dynamics \cite{Veneziano:1982ah,Ferrara:1982qs, Affleck:1983mk,Dine:1985rz}. This is a natural possibility, since non--abelian hidden gauge theories are generally present in heterotic string constructions.
An interesting mechanism of this sort is the so--called ``racetrack" mechanism, where one assumes  the condensation of several non--abelian gauge theories leading to a superpotential admitting non--trivial
minima for the moduli \cite{Krasnikov:1987jj}.\footnote{Another interesting possibility is to assume the presence of certain radiative corrections for the  K\"ahler potential of the moduli \cite{Banks:1994sg}. In absence of a calculable expression for such corrections, however, this mechanism is 
less predictive and, moreover, still needs the presence of non--perturbatively induced superpotentials.}
The racetrack mechanism has indeed been applied with success in the past to stabilize the dilaton field $S$ and other moduli as well, such as the universal K\"ahler modulus $T$,  see e.g. \cite{Casas:1990qi,deCarlos:1992da} for applications in the context of heterotic string theory.
One of the main drawbacks encountered in the past to stabilize moduli using the racetrack mechanism
was the need of large hidden sector gauge groups, often beyond the bound imposed by perturbative $E_8\times E_8^\prime$ or $Spin(32)/\Z_2$  heterotic strings. Another severe problem was  the difficulty in obtaining a de\,Sitter (dS) or Minkowski vacuum, since the resulting vacua turned out to be deep Anti de\,Sitter (AdS) vacua.

The main aim of this paper is to solve the above problems and to show that it is possible to stabilize the dilaton $S$ and the universal K\"ahler modulus $T$ in a dS/Minkowski vacuum with low energy dynamical supersymmetry breaking~\cite{Witten:1981nf}, using non--perturbatively generated superpotentials for the moduli. The main new ingredient we add, in addition to the well--known results on gaugino condensation,  is the IR dynamics of super Yang--Mills theories with light flavours, recently analyzed by Intriligator, Seiberg and Shih (ISS)~\cite{Intriligator:2006dd}.\footnote{See e.g.~\cite{Franco:2006ht,Ooguri:2006bg,Braun:2006da,Bena:2006rg,Argurio:2007qk} 
for recent applications of the results of \cite{Intriligator:2006dd} in constructing string models with dynamical SUSY breaking in meta--stable vacua.}
We then study the dynamics of the moduli in presence of  non--perturbatively generated superpotentials for them, in interaction with  the mesons $\Phi$ and the baryons (or dual magnetic quarks) $\varphi$ and $\tilde \varphi$ of~\cite{Intriligator:2006dd}. 

All our analysis is performed at the supergravity level using an effective field theory approach. 
Several simplifying assumptions are made,
in order to restrict the theory to a tractable system.
First of all,  we focus our attention
on the dynamics of $S$ and $T$ only, neglecting all other moduli altogether.
We also assume that all D--terms (and the F--terms for most of the charged fields) vanish and that the resulting theory has only (exotic) vector--like charged fields and several non--abelian unbroken Super Yang--Mills (SYM) theories  in the hidden sector. After integrating out all massive vector--like matter fields, we are left with low energy effective strong coupling scales for the pure SYM theories. The only exception is provided by the flavours of the ISS sector, which are taken to be massless at high energies and supposed to get a light mass by some dynamical mechanism, based on some of the hidden sector gauge groups present in the model.  We also assume that the underlying string model does not have an anomalous $U(1)_X$ gauge field, so that all gauge and gravitational anomalies are cancelled without need of any Green--Schwarz mechanism~\cite{Green:1984sg,Dine:1987xk}. 
All our interest will be in the hidden sector of the theory, where moduli stabilization and dynamical SUSY breaking is supposed to occur. We will not discuss how the SUSY breaking is mediated to the visible sector, which is assumed to be an SU(5) or SO(10) GUT theory, possibly broken to the Standard Model gauge group by Wilson lines. We only mention that gravity mediation of SUSY breaking seems preferred
to avoid very light moduli, linked to the gravitino mass in our framework, 
as they would be cosmologically problematic~\cite{deCarlos:1993jw,Banks:1993en}.\footnote{The ``overshooting problem''  \cite{Brustein:1992nk} --- another common
cosmological problem when stabilizing moduli with racetrack potentials ---  seems less severe;  see e.g. \cite{Barreiro:1998aj}.}  Lowering the gravitino mass scale below, say, 100 GeV also requires more
severe fine--tuning in the model and/or leaving the perturbative regime of the heterotic string which, 
in terms of $S$ and $T$, is essentially the requirement $S,T\gtrsim M_P$, where $M_P$ is the (reduced)
Planck scale.

The superpotential of the theory consists of two parts: one coming from the condensation of the pure SYM theories and another arising from an ISS--like model with $N_f=N_c+1$ flavours.
For simplicity, we call the two parts respectively as the racetrack and ISS sectors.
They are coupled, not only due to gravity, but also due to the universal nature of all gauge kinetic functions in heterotic string theory, determined by $S$ at tree--level.
One of the crucial points of our analysis is to show that, despite this non--decoupling, 
one can study the system and reliably control the back-reaction of one sector with respect to the other.
Roughly speaking, the racetrack sector is mainly responsible for the stabilization of the moduli, 
whereas the ISS sector provides the main source of SUSY breaking with a large $F$--term
in the meson direction, $F_\Phi>F_T\gg F_S$, needed to get a dS/Minkowski vacuum.
 
On more physical grounds, we can summarize the dynamics of moduli stabilization 
in our models as follows. At some energy scale $\Lambda$, the strong coupling dynamics of two or three
SYM theories in the racetrack sector stabilizes $S$ and $T$ in an AdS minimum. The moduli have Planckian VEV's
but are light, with a mass parametrically given by $\Lambda^3/M_P^2$. 
At a scale $ \Lambda_{ISS}< \Lambda$, but not $\ll \Lambda$, 
the ISS dynamics enters into the game.
Since $\Lambda^3/M_P^2\ll \Lambda_{ISS}$, the moduli cannot be integrated out and have to be retained in the effective theory. However, due to their tiny gravitational interactions with matter,
they provide negligible corrections to the ISS dynamics as analyzed in~\cite{Intriligator:2006dd}.
At low energy the ISS sector provides another non--perturbatively generated
superpotential term for the moduli. This causes a slight shift on the VEV's of $S$ and $T$, but
most importantly it can ``uplift'' the previous AdS minimum to a dS/Minkowski one. As in~\cite{Intriligator:2006dd},
the minimum turns out to be only meta--stable. It can decay into the SUSY Minkowski run--away minimum at $S\to\infty$ and into a deeper nearby AdS SUSY vacuum. The latter decay rate is by far larger than the former, yet it is small enough to allow for a cosmologically  long--lived vacuum.

There are three essential scales in the theory. The Planck scale $M_P$, which is the scale
of the VEV's of $S$ and $T$,  an intermediate scale $\mu\sim {\cal O}(10^{11})$ GeV  which fixes the mass scale of the ISS mesons and baryons and a light scale $\mu^2/M_P$, which is the mass scale of the moduli and of the gravitino.  For concreteness, we will focus our attention to two particular classes of racetrack superpotentials, based on two or three condensing gauge groups. We call the corresponding models RT2 and RT3. In the RT2 model, supersymmetry is broken at the minimum even in the absence of the ISS sector,
whereas in the RT3 model the minimum is SUSY in this limit. In both models the dilaton acquires
a mass ${\cal O}(10^3\, m_{3/2})$, with $m_{3/2}$  the gravitino mass, whereas the K\"ahler modulus is lighter, ${\cal O}( 10\,  m_{3/2})$ in the RT3 model and ${\cal O}(m_{3/2})$ in the RT2 one. 
We report various quantities of interest for two particularly promising models in Table~\ref{tab.1}.

The paper is organized as follows. In section~\ref{racetracksect} a brief review of the racetrack mechanism is given,
the RT2 and RT3 models are presented, and a simple theory consisting of a racetrack sector
coupled to an O'Raifeartaigh--like model is analyzed. The resulting toy model is useful
to understand in a simple set--up several features of the more realistic models subsequently constructed. In section~\ref{realisticmodels}, these realistic models are considered. After a brief review of the results of~\cite{Intriligator:2006dd}, we analyze in some detail the non--SUSY and SUSY vacua obtained by coupling a racetrack sector
with an ISS sector. In section~\ref{general}, some generalizations and further possible constraints arising from a string embedding of such models are discussed. Section~\ref{numerics} is devoted to a numerical analysis
of the two specific classes of models, RT2  and RT3, whose essential results are reported in Table~\ref{tab.1}.
In section~\ref{decay} we estimate the life--time of the meta--stable non--SUSY vacua and show that they are long--lived enough. Finally, some conclusions are given in section~\ref{concl}.

\section{The Racetrack Mechanism}\label{racetracksect}

The so--called racetrack is a mechanism to stabilize the chiral fields 
governing the holomorphic gauge kinetic terms of a supersymmetric theory with two or more non--abelian gauge 
groups~\cite{Krasnikov:1987jj}. In most cases of interest, supersymmetry and Peccei--Quinn like shift symmetries forbid  any perturbative superpotential term for these fields \cite{Witten:1985bz}, which are then chiral moduli fields to all orders in perturbation theory.
At low energies, the non--abelian gauge groups undergo gaugino condensation and the resulting non--perturbatively generated superpotential consists of two or more terms whose competing effects lead to a stabilization of the moduli. Before entering into some details of the racetrack mechanism, let us recall
that in perturbative heterotic string theory (with the assumption of supersymmetric grand unification), the string scale, the compactification scale and the reduced Planck scale are tied to be of the same order of magnitude \cite{Witten:1996mz}.  This implies $S_0\sim T_0\gtrsim M_P$,  in terms of the VEV's of the low--energy fields $S$ and $T$.  More precisely, one should require  ${\rm Re}\, S_0\sim 2$ at the field--theory GUT scale $2\times 10^{16}$ GeV.
It is well known that the heterotic GUT scale  (which is essentially identified with the string scale) is typically more than one order of magnitude higher than the field--theory GUT scale. For simplicity, in this paper, we will not enter in these issues and identify the string scale $M_s$ and the compactification scale $M_c$ (defined as the mass of the first Kaluza--Klein vector resonance) with the reduced Planck scale $M_P=2.4\times 10^{18}$ GeV, as the only UV scale in the theory. In terms of the moduli $S$ and $T$, one has approximately $M_c \simeq M_s/\sqrt{{\rm Re}\,S\,{\rm Re}\,T}$. Heterotic strings are on the edge of perturbativity in both the string coupling
and $\al$ expansions.

\subsection{Single Modulus Case}\label{singmod}

For a single SYM theory with holomorphic gauge kinetic term of the form\footnote{We follow the conventions of \cite{WessBagger}, use units in which $M_P=1$  and normalize the generators so that ${\rm \,Tr} \,t_a t_b = \delta_{ab}$ in the fundamental representation.  For simplicity,   we consider only level one Kac--Moody  groups.}
\be
f(S) {\rm Tr}\, W^2 =\frac{S}{4} {\rm Tr}\, W_\alpha W^\alpha +c.c. \supset  -\frac 14 {\rm Re \, S}\,  F_{\mu\nu}^a F^{\mu\nu,a} -\frac18{\rm Im \,S}\, \epsilon_{\mu\nu\rho\sigma} F^{\mu \nu,a} F^{\rho\sigma,a} \,,
\label{SWcoup}
\ee
where $S$ is the modulus (dilaton) field, a symmetry argument \cite{Dine:1985rz}\footnote{Arguments based on the Veneziano--Yankielowicz  superpotential \cite{Veneziano:1982ah} give the same functional form.} allows to fix the form of the non--perturbatively generated superpotential resulting from the condensation of the non-abelian gauge group. One gets
\be
W(S)= A e^{-\frac{24 \pi^2  S}{b}}\,,
\label{gaugino-single}
\ee
where $b$ is the coefficient of the one--loop beta function: $\beta(g) =- b\, g^3/16 \pi^2$,
 $A$ is a constant to be determined and $S$ is the dilaton value at the reduced Planck scale. One easily recognizes that the scalar component of eq.~(\ref{gaugino-single}) is proportional to $\Lambda^3$, where $\Lambda$ is the dynamically generated scale, as expected.
In presence of several condensing gauge groups, no simple symmetry argument allows to fix the form
of the effective superpotential for $S$. In absence of (light) charged matter, it is natural to assume that $W(S)$ will be the simple sum of the various
non--perturbatively generated superpotentials:
\be
W(S) = \sum_i A_i e^{-\frac{24 \pi^2 S}{b_i}}\,.
\label{gaugino-multiple}
\ee
Indeed, the only couplings between the gauge sectors are mediated by gravity and $S$, but
both interactions are too weak to modify considerably the much stronger gauge dynamics.

{}For  two gauge groups with one-loop coefficients $b_i\ll 24\pi^2$ --- so that  the coefficients multiplying $S$ in the exponents in $W(S)$ have coefficients much larger than 1 ---
it is straightforward to get a good analytical approximation for the dilaton VEV $S_0$ and its mass.
The usual supergravity (SUGRA) scalar potential in absence of D--terms is \cite{Cremmer:1982en}
\be
V=e^{K} (K^{S\bar S} D_S W \overline{D_{S}W} -3 |W|^2)\,,
\ee
where 
\be
K=-\ln(S+\bar S)
\label{KStree}
\ee
is the tree--level K\"ahler potential for $S$ \cite{Witten:1985xb} and 
$D_S W=\partial_S W +(\partial_S K) W$. 
Modulo accidental cancellations, $\partial^2_{S} W \gg  \partial_S W \gg W$ and the condition of extremum for $V$ is approximately given by the cancellation
of the leading $\partial_S^2 W$ term, implying either
\be
i) \ \ \partial_S W\simeq  D_S W =0 \ \ \ {\rm or}  \ \ ii) \ \  \partial^2_S W = 0\,.
\label{extrema}
\ee
The extremum $i)$ corresponds to a minimum of the scalar potential whereas $ii)$ is a minimum along the axion direction and a maximum along the
real dilaton direction, i.e., it is a saddle point. The extremum ii) corresponds to the barrier separating the minimum i) from the run--away minimum at infinity. The condition $i)$ in eq.~(\ref{extrema}) is solved for
 \be
 S_0 \simeq \frac{b_1 b_2}{24\pi^2(b_1-b_2)} \ln\Big(-\frac{A_2}{b_2}\frac{b_1}{A_1}\Big)\,.
 \label{S0vev}
 \ee
 It is trivial to verify from eq.~(\ref{S0vev}) that the axion VEV ${\rm Im}\,S_0$ is always such that
 the two coefficients at the extremum have opposite coefficients, so it is not restrictive to take $A_1$ and $A_2$ real, with $A_1>0$ and $A_2<0$. For $b_1\sim b_2\ll 24 \pi^2$ and $|A_1|\sim |A_2|$, eq.~(\ref{S0vev}) implies $S_0\ll 1$.
 Some moderate tuning between the $b_i$'s and/or the $A_i$'s
is then required to get $S_0\sim 1$. Since $F_S=D_S W= 0$, SUSY is unbroken and the physical mass for the whole dilaton multiplet is approximately given by 
  \bea
 m^2_S & \simeq & 8 S_0^3 \partial^2_S W(S_0) = 8 a_1^2 A_1^2 (a_1-a_2)^2 S_0^3 e^{-2 a_1 S_0}\nn \\
 &  = & \frac{ 8 a_1^2 A_1^2}{a_1-a_2} \left(\frac{a_1 A_1}{a_2 |A_2|}\right)^{-\frac{2 a_1}{a_1-a_2}} \ln^3\Big(\frac{A_1 a_1}{|A_2| a_2}\Big) \,,
 \label{m2dilaton}
  \eea
 where for convenience we have defined $a_i\equiv 24\pi^2/b_i= 8\pi^2/N_i$, with the last equality 
 valid for pure $SU(N)$ SYM theories.  Eq.~(\ref{m2dilaton}) is valid for $a_1>a_2$ and $|A_1| a_1 >  |A_2| a_2$. 

Let us now give an estimate of the expected values of the coefficients $A_i$ appearing
in eq.~(\ref{gaugino-multiple}).  As long as the holomorphic gauge kinetic functions are well approximated by their tree-level value $S/4$, it is reasonable to expect that the $A_i$ should not differ much from their flat-space value, in absence of any modulus. In the simple case of a pure $SU(N_c)$ gauge group, for instance, one has $A=N_c$, so that
\be
W(S)  = N_c \Lambda^3 =  A e^{-a S}, \ \ {\rm with} \ \ \  |\Lambda|   =  e^{-\frac{8\pi^2}{g^2 3N_c}}\,.
\label{flatEffPot}
\ee
In presence of massless matter, eq.(\ref{flatEffPot}) is replaced by an Affleck-Dine-Seiberg kind of superpotential  \cite{Affleck:1983mk}. 
 For simplicity, we will assume in the following that no light charged matter is present,\footnote{See 
\cite{Casas:1990qi,deCarlos:1992da}  for studies of SUSY breaking and moduli stabilization in heterotic string--inspired theories in presence of several non--perturbatively generated superpotentials and charged matter.} but we will allow for the possibility of having in the microscopic theory some massive charged matter with a mass $m < 1$, but not $m\ll 1$, for naturalness reasons. This mass might arise from a trilinear coupling in the superpotential (say, with some other gauge singlet modulus),  from a non--renormalizable coupling with charged fields with VEV along D-flat directions,  be a relatively light Kaluza--Klein excitation of a slightly anisotropic Calabi--Yau compactification, etc. In the following,
we will simply assume the presence of these massive charged states without any further investigation
of their dynamics. Once integrated out, the effects of such states is to give rise to $m$--dependent coefficients: $A_i=A_i(m)$. In the simple case of 
$N_f$ pairs of $SU(N_c)$  fundamentals and anti--fundamentals with a common mass  $m$, one gets (see e.g. \cite{Intriligator:1995au})
\be
W(S) = N_c \Lambda^{3} m^{N_f/N_c} =  A e^{-a S}\,, 
\label{effectiveA0}
\ee
with $\Lambda$ and $a$ as in eq.(\ref{flatEffPot}) and
\be
A= N_c m^{N_f/N_c}
\label{effectiveA}
\ee
a mass--dependent coefficient.
Eq.(\ref{effectiveA}) is actually also valid for $N_c$ quark fields in the fundamental of $SO(N_c+2)$.
In presence of a sufficient number of flavours,\footnote{For massive flavours, eq.~(\ref{effectiveA}) makes sense for any $N_f$.} eq.~(\ref{effectiveA}) implies that $A$ can naturally be of a few orders
of magnitude lower than  the flat pure SYM value $N_c$.\footnote{Moduli--independent threshold corrections to the gauge kinetic functions might also provide a displacement of $A$ from its ``standard'' value $N_c$, but generally no more than by a factor of order 1.}  In this way, one can easily check that a mild tuning between
$A_1$ and $A_2$ in eq.~(\ref{S0vev}) would allow to get $S_0\gtrsim 1$ with sufficiently low rank gauge groups to be  accommodated in perturbative $Spin(32)/\Z_2$ or $E_8\times E_8^\prime$ string models.

$S$ being stabilized at a SUSY point, we necessarily get a large negative cosmological constant, of order $m_{3/2}^2$, where $m_{3/2}^2$ is the gravitino mass term in the supergravity action.\footnote{It has been shown in \cite{Casas:1996zi} that no global minima with $S\simeq 2$ and vanishing cosmological constant exist by taking the tree-level K\"ahler potential (\ref{KStree}) and an arbitrary superpotential for S. Local minima are in principle possible, but they require severe tunings in the superpotential. More recently, it has been shown, using the K\"ahler potential (\ref{KStree}), 
that the scenario where SUSY is broken mainly in the $S$ direction  
is not compatible with the requirement of a meta--stable SUGRA vacuum \cite{GomezReino:2006dk}.}
 The stabilization of a modulus by means of a racetrack requires then some extra sector up--lifting the AdS vacuum to a dS/Minkowski one.\footnote{Of course, the requirement of vanishing cosmological constant applies to the physical vacuum energy and not at its tree--level value. An up--lifting sector is required 
if the latter is bigger than the typical  one--loop correction ${\cal O}(m_{3/2}^2/(16\pi^2))$, as in the case under consideration. }
 If the up--lifting sector does not provide for a large change on the value of the superpotential at the minimum, the gravitino mass is approximately given by 
\be
m_{3/2}^2 \simeq e^{K(S_0)} |W_0|^2 = \frac{A_1^2 (a_1-a_2)^2}{a_2^2 2S_0}e^{-2a_1 S_0}
=  \frac{A_1^2 (a_1-a_2)^3}{2a_2^2 \ln\Big(\frac{A_1 a_1}{|A_2| a_2}\Big)}
 \left(\frac{a_1 A_1}{a_2 |A_2|}\right)^{-\frac{2 a_1}{a_1-a_2}}  \,.
\label{gravitinomass}
\ee 
Eq.s~(\ref{m2dilaton}) and (\ref{gravitinomass}) give
 \be
 \frac{m_S}{m_{3/2} }\simeq 4 S_0^2 a_1 a_2 \gg 1\,,
  \ee
 implying a dilaton mass considerably higher than the scale of SUSY breaking fixed by $m_{3/2}$.
 
  \subsection{Two Moduli Case}

In heterotic string models, in addition to constant corrections to the gauge couplings, 
sizable moduli--dependent threshold corrections can appear, depending on the underlying string construction. When this happens, the gauge couplings, governed at tree--level by $S$ only, become function of many extra moduli, including the universal K\"ahler modulus $T$. In particular, this feature may be used to stabilize several untwisted moduli in heterotic string theory. 
Focusing only on the field $T$, the holomorphic gauge kinetic functions  read 
\be
4 f(S,T)= S + \epsilon \ln \eta(iT)\,,
\label{holoGKF}  
\ee
where $\eta(i T)$ is the Dedekind eta function 
(conventions as in \cite{polV1}) and the $\ln \eta(iT)$ term is the well known holomorphic 
moduli--dependent threshold correction term \cite{Dixon:1990pc}, which in general depends on the gauge group. The coefficient $\epsilon$ is one--loop induced, and roughly goes like $1/(8\pi^2)$ times a group theoretical factor, proportional to the $\beta$--function of an ${\cal N}=2$ SUSY theory associated to the original ${\cal N}=1$ one (for details, we refer to \cite{Dixon:1990pc}). 
The non--perturbatively generated superpotential (\ref{gaugino-single}) becomes now
\be
W(S,T) = A e^{-a (S+\epsilon \ln\eta(i T) )}\simeq A e^{-a (S - \pi T \epsilon/12)}
\label{gaugino-ST}
\ee
where the last approximation is valid for $T\geq 1$. Since $\epsilon$ is model--dependent,
we can define a phenomenological parameter $\gamma\equiv -a \pi \epsilon/12$, whose typical size 
is in the range $-1\lesssim \gamma\lesssim 1$. For simplicity, we focus our attention to superpotential terms of the form
\be
W_{RT}(S,T)=W(S) e^{-\gamma T} - \hat W(S)\,,
\label{SPotRT2RT3}
\ee
where 
\be
\hat W(S) = A_1 e^{-a_1 S}\,,
\ee
and $W(S)$ is either a simple exponential or a sum of two of them.
We denote by  ``RT2'' and ``RT3'' the respective models. In order to be able to provide some approximate analytical formulae for the VEV's of $S$ and $T$, as in subsection 2.1, we take $W^{\prime\prime} \gg   W^\prime \gg W$ 
and $\hat W^{\prime\prime} \gg  \hat W^{\prime} \gg \hat W$, where a prime stands for a derivative with respect to $S$.

\subsubsection{The RT2 Model} 

This is defined by taking
\be
W(S)=A_2 e^{-a_2 S}
\label{rt2Def}
\ee
in eq.(\ref{SPotRT2RT3}).
 Let us start by looking for supersymmetric configurations. The condition $F_S=0$
at leading order gives 
\be
W^\prime = \hat W^\prime e^{\gamma T} \,.
\label{FSvanish}
\ee
Substituting eq.~(\ref{FSvanish}) in the $F_T=0$ condition gives, after trivial algebra,
the approximate VEV's for $S$ and $T$  (see \cite{Abe:2005pi} for a similar analysis):
\bea
T_{\rm SUSY} & & \simeq \frac{a_2-a_1}{a_1} \frac{3}{2\gamma}\,, \nn \\
S_{\rm SUSY} & & \simeq \frac{1}{a_2-a_1} \ln\Big(\frac{a_2A_2}{a_1A_1}\Big) -\frac{3}{2a_1}\,.
\label{STvevSUSYRT2}
\eea
It turns out, however, that the extremum (\ref{STvevSUSYRT2}) corresponds to an AdS 
saddle point in the field region of interest ($S,T\sim 1$). After uplifting, this vacuum will typically
give rise to tachyonic directions. For this reason, we now look for non--supersymmetric minima.
The potential is extremized, at leading order, by 
\bea
W^{\prime\prime} & = &  \hat W^{\prime\prime} e^{\gamma T} \label{extremaRT2} \,, \nn \\
W^{\prime}&  = &  \hat W^{\prime}\frac{e^{\gamma T}}{1+2/3 \gamma T} \,.
\eea
whose solutions are given by
\bea
T_0  & & \simeq \frac{a_2-a_1}{a_1} \frac{3}{2\gamma}\,, \nn \\
S_0 & & \simeq \frac{1}{a_2-a_1} \ln\Big(\frac{a_2^2A_2}{a_1^2A_1}\Big) -\frac{3}{2a_1}\,.
\label{STvevNOSUSYRT2}
\eea
The extremum (\ref{STvevNOSUSYRT2}) is a minimum of the potential. The largest term in the scalar mass matrix for the scalars
is $V_{S\bar S}$, which then fixes the (physical) mass for the dilaton to be given by
\be
m_S^2\simeq  \frac{S_0^3}{T_0^3}  W^{\prime\prime}_{RT}(S_0,T_0)^2\,.
\label{mSRT2}
\ee
There is no a similar,  simple and accurate enough formula for the mass of $T$ because the off--diagonal terms $V_{S\bar T}$ and $V_{\bar ST}$ in the mass matrix cannot be neglected.  It is nevertheless possible to see that its mass is at most  ${\cal O} (W^\prime_{RT})$ and hence lighter than $S$ by a factor 
$a_{1,2}$. In fact, as verified by a numerical analysis (see e.g. Table~\ref{tab.1}), the mass of the K\"ahler modulus is typically three orders of magnitude smaller
than that of the dilaton.

\subsubsection{The RT3 Model}

This is defined by taking
\be
W(S)=A_2 e^{-a_2 S}-A_3 e^{-a_3 S}
\label{RT3}
\ee
in eq.(\ref{SPotRT2RT3}). Contrary to the RT2 model, the RT3 model admits supersymmetric minima, so that we focus here on supersymmetric configurations only.  Due to the presence of two exponentials
in eq.~(\ref{RT3}), it is not possible to write, as before, analytical formulae for $S$ and $T$ accurate enough, in general.
However, if  $\Lambda_2\simeq \Lambda_3>\Lambda_1$ but not $\gg \Lambda_1$, it is possible to disentangle the $S$ and $T$ stabilization
from each other.  Indeed, in such a case, the VEV of $S$ is to a very good approximation determined entirely by $W(S)$. Once fixed $S$, the 
superpotential (\ref{SPotRT2RT3})  resembles the KKLT superpotential of type IIB flux compactifications and stabilizes $T$ \cite{KKLT}. 
As before, the gravitational terms in $F_S$ are subleading, so that 
\be
S_0\simeq \frac{1}{a_2-a_3} \ln\Big(\frac{A_2 a_2}{A_3 a_3}\Big)\,.
\ee
In  $F_T$ the gravitational terms are important and one finds that $F_T\simeq 0$ for
\be 
T_0\sim -\frac{1}{\gamma}\ln\bigg(\frac{\hat W(S_0)}{W(S_0)}\bigg)\,.
\label{tVEV}
\ee
The mass for the dilaton is approximately given by
\be
m_S^2\simeq  \frac{S_0^3}{T_0^3} e^{-2\gamma T_0}W^{\prime\prime}(S_0)^2\,,
\label{mSRT3}
\ee
whereas the K\"ahler modulus is again much lighter.

Finally, note that in both the RT2 and RT3 models the minus signs in the superpotential were chosen such that  the minima of $S$ and $T$ are real, with no
VEV's  for the axionic components.

Just to give an idea of the amount of accuracy of the above analytical formulae, we compare here the approximate analytical values for $S_0$, $T_0$ and $m_S$
to those given in the Table~\ref{tab.0}, obtained numerically from the full potential. The input parameters $A_{1,2,3}$, $a_{1,2,3}$, $\gamma$ and $A_{1,2}$, $a_{1,2}$, $\gamma$ 
entering in eq.s~(\ref{SPotRT2RT3})--(\ref{rt2Def}) and (\ref{RT3})
for the RT3 and RT2 models  have been chosen as given in Table~\ref{tab.1}.

\TABLE[t]{
\begin{tabular}{| l | l | l |}
\hline &
$\textrm{analytical}$ & $\textrm{numerical}$ \\
& $\textrm{approximation}$ & $\textrm{solution}$ \\
\hline 
$\textrm{RT3}$:  &  \hspace{1ex}$S_0\sim1.26$ & \hspace{1ex}$S_0\approx 1.18$ \\
 &  \hspace{1ex}$T_0\sim1.39$ & \hspace{1ex}$T_0\approx 1.55 $ \\
 &  $m_{S}\sim1.1\cdot 10^6\,{\rm GeV}$ & $m_{S}\approx 3.3\cdot 10^6\,{\rm GeV}$ \\
 \hline
$\textrm{RT2}$: &  \hspace{1ex}$S_0\sim1.90$ & \hspace{1ex}$S_0\approx 1.63$ \\
 &  \hspace{1ex}$T_0\sim2.50$ & \hspace{1ex}$T_0\approx 3.3$ \\
 &  $m_{S}\sim1.0\cdot 10^5\,{\rm GeV}$ & $m_{S}\approx 2.0\cdot 10^6\,{\rm GeV}$ \\
 \hline
\end{tabular}
\caption{Moduli VEV's, and dilaton masses for the local minima in the two racetrack setups RT2 and RT3 in the analytical approximation (left column) and the full numerical solution from the racetrack scalar potential (right column).}
 \label{tab.0}
}

As can be seen from Table~\ref{tab.0}, the values of $S$ and $T$ obtained analytically do not differ too much from the actual numerical values, whereas the masses
differ by factors of order one or more. This should not surprise, because of the exponential sensitivity of physical parameters to the VEV's of the moduli, typical
in racetrack models. Hence, care should be taken in using the analytical relations we found above.

\subsection{A Toy Model for a dS/Minkowski vacuum}\label{toy}

We have seen in the last section that the simplest model of dilaton stabilization by means of two  condensing gauge groups leads to a SUSY AdS vacuum. Upon the inclusion of threshold corrections $T$ gets stabilized as well. SUSY can be broken (typically with $F_T\gg F_S$) or not, depending on whether the racetrack superpotential consists of 2 or 3 gauge groups (RT2 or RT3), respectively. Yet the vacuum remains a deep AdS vacuum. The addition of light charged matter does not seem
to change much the situation (see e.g. \cite{deCarlos:1992da}). Of course, one can invoke
any extra dynamics further breaking SUSY, in this way providing the additional F or D--terms needed to cancel the cosmological constant. In order not to interfere too much with the moduli
stabilization mechanism, one typically asks for an extra sector which  does not provide too large a back-reaction for the dynamics stabilizing the moduli, so that the latter fields can essentially
be taken as frozen at their VEV's. Decoupling the moduli stabilization dynamics from the main source of SUSY breaking can also allow for moduli masses much higher than the SUSY breaking scale,
a certainly welcome feature for cosmological reasons. Much work has been recently devoted to this
``up--lifting'' problem, mainly in the context of IIB compactifications, after the KKLT scenario proposed
in \cite{KKLT}. Since in heterotic string theory the dilaton couples universally to all gauge fields, it is not easy  to find an up--lifting sector which is completely decoupled
from the dilaton stabilization sector.  Nevertheless, as long as the dynamics in the latter 
is strong enough, the up--lifting sector will not alter much the dilaton stabilization mechanism, allowing
for a partial decoupling between the two sectors. 

The simplest ($F$--term) SUSY breaking that one can invoke is the O'Raifeartaigh  model 
\cite{O'Raifeartaigh:1975pr}, namely a simple non--generic polynomial superpotential which does not admit SUSY solutions. This model has recently had a renewed interest  after Intriligator, Seiberg and Shih (ISS) pointed out
that the IR dynamics of $SU(N_c)$ ${\cal N}=1$ SUSY gauge theories with $N_c < N_f < 3/2 N_c$
light pairs of quarks in the fundamental and anti--fundamental representations of $SU(N_c)$ (as well as
$SO(N_c)$ and $Sp(N_c)$ theories with appropriate matter)
admits meta--stable vacua where SUSY is broken \cite{Intriligator:2006dd}. The revival of the old proposal of \cite{O'Raifeartaigh:1975pr} arose because the effective superpotential describing the non--SUSY ISS vacua is essentially a sum of  O'Raifeartaigh -- like models. Since we are going to consider as the up--lifting sector a SYM theory with light flavours, along the lines of \cite{Intriligator:2006dd}, in the following we study in some detail a toy model consisting of an O'Raifeartaigh model coupled to a racetrack superpotential coming from two condensing gauge groups.\footnote{See \cite{OKKLT2} 
for a recent application of an O'Raifeartaigh model as an up--lifting sector in IIB SUGRA theories.}
We neglect gravitational interactions, which  do not play an important role, and study the model in flat space. Gravity will be eventually considered only to estimate the amount of SUSY breaking required to cancel the negative cosmological constant. This model is too simple to capture all the properties of the more refined model we present in the next section, but it illustrates its most important features.

The K\"ahler and superpotential are the following:
\bea
K & = &  -\ln(S+\bar S) +X^\dagger X + \phi_1^\dagger \phi_1+ \phi_2^\dagger \phi_2
\,, 
\label{KOraife} \\
W & = &  W_{RT}(S) + W_{OR}(S,X,\phi_{1,2}) \nn \\
&  = & A_1 e^{-a_2 S}-A_1 e^{-a_2 S} + m \phi_2 \phi_1
+h X \phi_1^2 - \mu^2(S) X \,, 
\label{WOraife}
\eea
where $\mu^2(S)$ is a dynamically generated scale, whose form is then fixed to be, up to a constant,
\be
\mu^2(S) = e^{-\eta S}\,,
\ee
with $a_1\simeq a_2$ and $a_{1,2}\lesssim \eta < 2 a_{1,2}$. The F--term vanishing conditions are satisfied only for $\phi_{1,2}=0$,  arbitrary $X$ and ${\rm Re}\, S = \infty$, corresponding to the usual trivial run--away minimum.
Any other minimum is then SUSY breaking and necessarily meta--stable. The obvious region in field space where to look for non--SUSY vacua is $\phi_{1,2} \simeq 0$ and
$S\simeq S_0$, with $S_0$ as in eq.~(\ref{S0vev}). 
Indeed, these are the VEV's for the fields when
the racetrack and the O'Raifeartaigh sector are decoupled. 
We take $m^2\gtrsim \mu^2(\tilde S_0)\gg m_S^2(\tilde S_0)$, 
with $\tilde S_0\simeq S_0$ the new minimum for $S$,
since this will be the typical parameter space occurring in our more realistic model (in which the mass $m$ is  $S$--dependent and dynamically generated as well). An important point to be stressed is the following.
In the model defined by eq.s~(\ref{KOraife}) and (\ref{WOraife}), the field $X$ is not a tree-level modulus, due to the interactions with the dilaton coming from the last term in eq.~(\ref{WOraife}). In particular, a quadratic term for  $|X^\dagger X|$  of order $\eta^2 \mu^4$ arises. One might conclude that $X$ will be stabilized at some given value due to the dilaton interactions, but this conclusion is in fact generally and obviously wrong, because it neglects one--loop corrections induced by the fields $\phi_{1,2}$ 
to $X$. The latter, although one--loop suppressed, are clearly much stronger than the tiny, gravitational in strength, interactions of $S$ with $X$. This is best seen if we actually integrate out $\phi_{1,2}$ and compute the resulting effective
K\"ahler potential for $X$ at one--loop level. This is straightforward, since one can safely neglect the
dilaton interactions so that the computation boils down to a standard O'Raifeartaigh model
with $\mu^2(S)$ fixed at $\mu^2(\tilde S_0)$.  For $m^2\gg \mu^2(\tilde S_0)$,
one gets
 \be
 K_{\rm eff}(X)  =  X^\dagger X 
 - \frac{h_R^4}  {12\pi^2 m^2} (X^\dagger X)^2
\,,
\label{effKahler}
\ee
where $h_R$ is the (renormalized) coupling defined as 
\be
\frac{d^4 V_{\rm eff}}{d^2\phi_1 d^2\bar \phi_1}\bigg|_{\phi_1\bar \phi_1=0} \equiv  4 h_R^2\,,
\ee
in terms of the effective tree+one--loop potential $V_{\rm eff}$.
The K\"ahler potential (\ref{effKahler}) gives rise to an $X^\dagger X$ term in the effective scalar potential of the kind
\be
K_{X^\dagger X}^{-1} |\partial_X W|^2 = \frac{h_R^4}{3\pi^2}\frac{\mu^4}{m^2}X^\dagger X + \ldots \,.
\label{Phi2K}
\ee
If $m^2$ is dynamically generated and proportional to $\mu^2$, it is clear that, despite  the loop factor suppression, the term in eq.~(\ref{Phi2K})  ${\cal O}(\mu^2)$ can (and actually will) be much larger than the tree-level value of order ${\cal O}(\mu^4)$ discussed above. Indeed, as we will see, $\mu$ turns out to be of order $10^{-7}$, so that it is necessary to include the one--loop correction above in the minimization of the potential arising from eq.s~(\ref{KOraife}) and (\ref{WOraife}). For all practical purposes, the effective K\"ahler potential (\ref{effKahler})  amounts to only adding the mass term (\ref{Phi2K}) for $|X|^2$ to the tree--level potential. It is now straightforward to see that the minimum
for $X$ is at the origin, as in the standard O'Raifeartaigh  model with no dilaton. 
The slight displacement of the VEV of $S$  from its unperturbed value $S_0$ due to the $X \mu^2$ interaction leads to a SUSY breaking in the $S$ direction,
$F_S(\tilde S_0)\neq 0$, but this is sub--leading with respect to the major source of SUSY breaking given by $F_X = -\mu^2(\tilde S_0)$. The latter effectively provides for the up--lifting term we were looking for. The requirement of having a vacuum with nearly zero energy requires that 
$F_X\sim W$ at $S=\tilde S_0$ once we include back gravity. This implies, in the approximation $\tilde S_0\simeq S_0$ and
by using eq.~(\ref{gravitinomass}),
\be
|W|^2\simeq \frac{A_1^2 (a_1-a_2)^2}{a_2^2} e^{-2a_1 S_0} \simeq e^{-2 \eta S_0} = \mu^4(S_0)\,,
\ee
and hence $\eta\gtrsim a_1$. When $m$ is $S$--dependent as well, with $m(S) \simeq \mu(S)$, the effective K\"ahler potential (\ref{effKahler})  is more involved and $S$--dependent as well, so that mixing between $X$ and $S$ is introduced.  Expanding for small $X$, one gets in $V_{\rm eff}$  mass terms for $X$ of the form $m^2$, $\mu^2$, $m^4/\mu^2$ times possible logarithmic terms. \footnote{In fact, the term $\mu^4/m^2$ in eq.(\ref{Phi2K}) arises from an expansion of such logarithmic terms.} As far as we focus on the region of $X$ close to the origin, however, the mixing between $X$ and $S$ is negligible and the only net effect of
having integrated $\phi_1$ and $\phi_2$ out is again a mass term for $X$, which differs from that in eq.(\ref{Phi2K}), but is still of the same order of magnitude.

There are essentially three relevant effective scales in the model: the Planck scale  which
sets the VEV taken by the dilaton, the dynamically generated scale $\mu=e^{-\eta S/2}\ll 1$, which is
the scale of mass for the O'Raifeartaigh fields $\phi_1$, $\phi_2$ and $X$, 
and the scale  $\mu^2 \ll \mu\ll 1$ which governs the dilaton and gravitino mass.
The requirement of a gravitino mass at the TeV scale fixes then the mass of the O'Raifeartaigh fields
in the typical regime of hidden sector models in gravity mediated SUSY breaking models 
\cite{GravityMed}, of order $10^{10\div 11}$ GeV. 

Summarizing, although the dilaton stabilization and up--lifting  sectors are not totally decoupled from each other, even in absence of gravity, nevertheless
the back-reaction of the latter on the former can be kept under control.

Let us conclude this section by noting that 
from a purely effective field theory point of view,  if we are interested
in the dynamics of the light fields,  we should integrate out 
the whole up--lifting sector, and study the resulting effective Lagrangian describing the dynamics of the
dilaton and of the gravitino only. Since the Goldstino (eventually eaten by the gravitino) is almost completely given by the fermion partner of $X$ or, in other words, since $F_X$ is the main source of SUSY breaking, the integration should necessarily be performed at the non--SUSY level.
Moreover, a one--loop (at least) integration is required when there are fields with tiny tree-level interactions  only, such as the field $X$  in the previous example. The whole integration procedure is then a bit involved, mainly when extended to the more realistic and complicated model of the next section. We have not followed such an effective approach in this paper
although it would be certainly interesting to do it, in particular to check the full quantum stability of the model.

\section{More Realistic Models}
\label{realisticmodels}

In this section we want to develop the toy model of moduli stabilization in presence of spontaneous supersymmetry breaking of the last section into a more realistic construction where the simple O'Raifeartaigh SUSY breaking sector is replaced by the IR sector of an ${\cal N}=1$ SYM model, along the lines of ISS \cite{Intriligator:2006dd} . 

\subsection{Brief Review of the ISS Model}

It was realized in  \cite{Intriligator:2006dd}  that the strong gauge dynamics of an ${\cal N}=1$ supersymmetric $SU(N_c)$ gauge theory with $N_{f}$ pairs of quark multiplets $Q$ and $\tilde Q$ in the fundamental and anti-fundamental representations of $SU(N_c)$ and mass matrix $m_f$  leads to meta-stable non--SUSY minima for $N_c< N_f<\frac{3}{2}N_c$.\footnote{See 
\cite{Dimopoulos:1997ww} for an earlier study of meta--stable non--SUSY minima in globally unbroken SUSY theories.} The vacua are parametrically long--lived if $\epsilon_{{\rm ISS}}\equiv \sqrt{m_f/\Lambda_{{\rm ISS}}}\ll 1$,  where $\Lambda_{{\rm ISS}}$ is the strong coupling scale of the SYM theory.  Below $\Lambda_{{\rm ISS}}$ the theory admits a perturbative description.
If  $N_c+1< N_f<\frac{3}{2}N_c$, this is given by an infrared--free ``magnetic theory'' with dual quark and meson fields of an $SU(N_f-N_c)$ SYM theory.
If $N_f=N_c+1$, a very similar perturbative description is given in terms of the baryons $\varphi \sim Q^{N_c}$ and $\tilde\varphi\sim \tilde Q^{N_c}$ and of the mesons $\Phi\sim \tilde Q Q$ of the original (``electric'') theory. In the following we focus on the case $N_f=N_c+1$. The dynamics of baryons and mesons at low energy is described by the superpotential \cite{Seiberg:1994bz}:
\be
W_{\rm ISS}(S,\varphi,\tilde \varphi,\Phi) =  {\rm Tr}\, \tilde\varphi^t \Phi\varphi  -  {\rm Tr}\, \mu^2 \,\Phi +\frac{{\rm det}\,\Phi}{\Lambda_{ISS}^{N_c-2}}\,, \label{SuperpotC0}\ee
in terms of (almost) canonically normalized fields with K\"ahler potential
\be
K_{\rm ISS}(\varphi,\tilde \varphi,\Phi)  =  {\rm Tr} \Big[\frac{1}{\alpha}\Phi^\dagger \Phi + \frac{1}{\beta}(\varphi^\dagger \varphi 
+ \tilde \varphi^\dagger \tilde\varphi)\Big]\, . \label{KahlerC}\ee
In eq.~(\ref{SuperpotC0}),
\be
 \mu^2 = m_f\, \Lambda_{\rm ISS}\label{mumf}\,,
\ee
and is taken real,  for simplicity. In the K\"ahler potential (\ref{KahlerC}), $\alpha$ and $\beta$  are incalculable coefficients assumed to be of order one. The baryons $\varphi$ and $\tilde \varphi$ are $N_c+1$ vectors, whereas the meson
$\Phi$ is an $(N_c+1)\times (N_c+1)$ matrix. It is convenient to parametrize these fields as follows: 
\be
\Phi  = 
\left(\begin{matrix} Y & Z^t \cr \tilde Z & \hat\Phi \cr \end{matrix}\right)\quad,\quad
\varphi = \left(\begin{array}{c} \chi \\ \rho \end{array}\right)\quad , \quad
\tilde\varphi = \left(\begin{array}{c} \tilde \chi \\ \tilde\rho \end{array}\right)\,,
\label{ISSfields}
\ee 
where $\chi$, $\tilde \chi$ and $Y$ are ordinary fields, $\hat\Phi$ is an $N_c\times N_c$ matrix and 
$Z$, $\tilde Z$, $\rho$ and $\tilde\rho$ are $N_c$ vectors. As shown in \cite{Intriligator:2006dd}, 
the above theory has a supersymmetry breaking vacuum near the origin in field space. In the simplest
case in which we take $m_{f,ij} = m_f \delta_{ij}$, the vacuum with maximal unbroken global symmetries
is given by 
\be
\langle \Phi \rangle \equiv \Phi_0  =  0 \,,  \ \ \ \ 
\langle \varphi \rangle \equiv  \varphi_0 = \left(\begin{array}{l}\mu \\ 0_{N_c} \end{array}\right)\,, \ \ \ \ 
 \langle \tilde \varphi \rangle \equiv  \tilde \varphi_0  = \left(\begin{array}{l}\mu \\ 0_{N_c} \end{array}\right)\,,
\label{ISSvacuum}
\ee
where $\mu\ll 1$. It is obvious that this vacuum is determined entirely by the first two terms in the superpotential \eqref{SuperpotC0}, since the determinant piece is negligible around the origin in $\Phi$.  The vacuum energy is given by $\mu^4$ and  the F-terms are $F_\varphi=F_{\tilde\varphi}=0$ and $F_\Phi\neq 0$ for the $N_c \times N_c$ components  $\hat \Phi\subset \Phi$. 
At the non-supersymmetric vacuum (\ref{ISSvacuum}), several fields acquire a tree--level mass 
${\cal O}(\mu)$, some  are Goldstone bosons of the broken global symmetries and remain massless at all orders in perturbation theory and the remaining ones are massless at tree-level only (pseudo--moduli) and acquire one--loop masses of order ${\cal O}( \mu/(4\pi))$. The pseudo--moduli fields are $\hat\Phi$ and ${\rm Re}\, (\chi-\tilde \chi)$. Around (\ref{SuperpotC0}), the theory has a structure which roughly resembles the O'Raifeartaigh like model introduced in eq.s~(\ref{KOraife}) and (\ref{WOraife}). Oversimplifying a bit, the field identifications are $X \rightarrow \hat \Phi$, $\varphi_1 \leftrightarrow \rho, \tilde \rho$, $\varphi_2 \leftrightarrow Z, \tilde Z$.

In addition to the non-supersymmetric vacuum (\ref{ISSvacuum}), the theory has  $N_c$ 
supersymmetric vacua for
\be 
\langle\varphi\rangle_{\rm SUSY}=\langle\tilde\varphi\rangle_{\rm SUSY}=0
\;\;,\;\; \langle\Phi\rangle_{\rm SUSY}=\frac{\mu}{\epsilon_{\rm ISS}^{(N_{c}-2)/N_c}}\, \unit_{N_{f}}\qquad 
.\label{susyvac1}
\ee
The non-supersymmetric vacuum  (\ref{ISSvacuum}) is then only metastable, but with a life--time
which is  parametrically long if
\be
\epsilon_{ISS}\equiv \frac{\mu}{\Lambda_{{\rm ISS}}}\ll1\,.
\label{epsilonISS}
\ee 
See \cite{Intriligator:2006dd} for more details.

The condition (\ref{epsilonISS}) requires an unnaturally small quark mass $m_f$. A possible way to overcome this tuning is to advocate a dynamical generation mechanism for $m_f$, along the lines of \cite{Dine:2006gm}. 
This can be done in various ways. For instance, in presence of one (or more) additional condensing gauge groups $G$ (as will be the case in our model, where they are responsible for the racetrack superpotential stabilizing $S$ and $T$), 
one might assume that the holomorphic gauge kinetic function for $G$
includes higher derivative operators of the form $Q\tilde Q/M^2 {\rm Tr}\,W^2$, where $W$ is the chiral field strength superfield associated to the gauge group $G$, and $M$ a high mass scale.
Alternatively, we can assume the presence of a quartic superpotential term coupling the ISS quarks $Q$ and $\tilde Q$ with quarks with a large mass $m$, e.g. like the ones advocated in subsection~\ref{singmod} to get the effective couplings (\ref{effectiveA}) in the racetrack superpotentials.
In this way, one has effectively the replacement $m\rightarrow m +  Q\tilde Q/M$ in eq.~(\ref{effectiveA}), giving rise to dynamically 
generated small mass terms $m_f$ for $Q \tilde Q$.
In the two cases one gets the following functional form of $m_f$ in terms of the strong coupling scale $\Lambda$
associated to $G$:
\bea
m_f & \sim & \frac{\Lambda^3}{M^2} \,,\nn \\  
m_f & \sim &   \left(\frac{\Lambda^3}{m}\right) \,.
\label{flavmasses}
\eea

\subsection{dS/Minkowski non--SUSY vacua with ISS}
\label{dS/M}

Along the lines of subsection~\ref{toy}, we consider here a scenario where two or more pure non--abelian gauge theories are responsible
for a moduli superpotential of the racetrack (RT) type, whereas a further $SU(N_c)$ gauge theory with $N_c+1$ light flavours gives rise to an effective superpotential for mesons and baryons as in eq.~(\ref{SuperpotC0}).
The model is defined by the straightforward sum of the K\"ahler and superpotential terms of the RT and ISS sectors:
\bea
K_{\rm tot} & = & K_{\rm RT}(S,T) + K_{\rm ISS}(S,T,\varphi,\tilde \varphi,\Phi), \label{Kahler} \\
W_{\rm tot} & = & W_{\rm RT}(S,T) + W_{\rm ISS}(S,\varphi,\tilde \varphi,\Phi)\,. \label{Superpot}
\eea
Here,  $K_{\rm RT} $ is the usual tree--level K\"ahler potential for the $S$ and $T$ 
moduli \cite{Witten:1985xb}
\be
K_{\rm RT}(S)  = -  \ln (S+\bar S) -3\ln (T+\bar T)
\label{Kahler-moduli}
\ee
and $W_{\rm RT}$ is the racetrack superpotential. We do not specify its form, because
it is not needed for the moment. The superpotential $W_{\rm ISS}$ has the same form as  in eq.~(\ref{SuperpotC0}), except
that now both $\Lambda_{\rm ISS}$ and $\mu^2$ are $S$--dependent. We do not know its explicit detailed form, but holomorphy 
and the axion symmetry of ${\rm Im}\,S$ broken only by non--perturbative effects essentially fix the functional form of the $S$--dependence entering in $W_{\rm ISS}(S)$ to be of exponential type. Modulo numerical coefficients, the form of $W_{\rm ISS}(S)$ 
which matches eq.~(\ref{SuperpotC0}) in the flat limit and dilaton decoupling is the following:
\be
W_{\rm ISS}(S,\varphi,\tilde \varphi,\Phi) =   {\rm Tr}\, \tilde\varphi^t \Phi \varphi- \, \mu^2(S)\, {\rm Tr} \,\Phi +e^{-8\pi^2 S\frac{2-N_c}{2N_c-1}} {\rm det}\,\Phi \,.
\label{SuperpotC}
\ee
Since we require $m_f$ to be dynamically generated as in eq.~(\ref{flavmasses}), the $S$--dependence of $\mu^2$ is not uniquely
fixed by $\Lambda_{\rm ISS}$. Hence, we write 
\be
\mu^2(S)=e^{-\eta S}\,,
\label{gammaDef}
\ee
where $\eta\gtrsim 16\pi^2/(2N_c-1)$ to fulfill the constraint (\ref{epsilonISS}).  
The ISS K\"ahler potential term in eq. (\ref{Kahler}) is the most uncertain term in our model, since there is no way to compute or argue in a reliable way its $S$ and $T$--dependence. As we will shortly see, however, in the field region of interest close to the non--supersymmetric ISS vacuum
(\ref{ISSvacuum}), all the terms in $K_{\rm ISS}$ are too small to affect the stabilization of $S$ and $T$, for a wide range of possible $S$ and $T$ moduli dependences entering in $K_{\rm ISS}$. It is then reasonable to freeze $S$ ad $T$ at their minima in $K_{\rm ISS}$ and recover the 
(almost) canonical K\"ahler potential eq.  (\ref{KahlerC}) modulo constant field redefinitions. We then take 
$K_{\rm ISS}(S,T,\varphi,\tilde \varphi,\Phi)$ to be independent of $S$ and $T$ and read as in eq.~(\ref{KahlerC}). 
We will relax this strong assumption in the following, showing how, in fact, no dramatic effect will result regarding the stabilization of $S$ and $T$.

As usual, the total scalar potential of the above SUGRA model, in the absence of D--terms, reads
\be
V_{\rm tot} = e^{K_{\rm tot}} \Big(K^{I\bar J}_{\rm tot}D_I W_{\rm tot} \overline{D_{J} W}_{\hspace{-0.3ex}\rm tot}  -3|W_{\rm tot}|^2\Big)\,,
\label{fullPot}
\ee
where $I,J$ run over all the chiral multiplets of the theory and 
$D_I W = \partial_I W + (\partial_I K) W$
is the K\"ahler covariant derivative.

\subsubsection{The meta-stable non-SUSY vacua}\label{nonSUSYvac}

An analytical study of the extrema of $V_{\rm tot}$ in eq.~(\ref{fullPot}) is a formidable task.
It is instead wiser to estimate the size of the many terms in $V_{\rm tot}$
in the field region of interest and hence decompose the total scalar potential (\ref{fullPot}) in a stronger
and a weaker component as
\be
V_{\rm tot} =  V_s +V_w \,, 
\label{potdecomposition}
\ee
with $|V_s|\gg |V_w|$.  Similarly to the toy model discussed in subsection (\ref{toy}), one has to be careful in identifying 
the relevant dynamical effects. It would not make sense to study the tiny gravitational and moduli corrections to the ISS vacuum 
without taking into account the much stronger radiative effects induced in the non--SUSY vacuum by the ISS fields themselves.
The latter are effectively included by adding a mass term for the pseudo--modulus $\hat \Phi$ in $V_{\rm tot}$ (see~\cite{Intriligator:2006dd}).\footnote{We do
not consider a similar mass term for ${\rm Re}\,(\chi-\tilde\chi)$ since it is not relevant in our analysis. Strictly speaking, the
added mass term for $\hat\Phi$ should be seen as coming from a one--loop correction to the K\"ahler potential, as in eq.~(\ref{effKahler}).
Since now $m\simeq \mu$, the form of  the K\"ahler corrected potential is more involved and not very enlightening. 
Around the vacuum (\ref{ISSvacuum}), its only relevant effect is of producing a mass term for $\hat\Phi$ in the scalar potential.}
The racetrack sector (plus the $\mu^4(S)$ term, see below)
leads to the stabilization of $S$ and $T$ at some values $S_0,T_0\gtrsim 1$, with $\mu_0^2\equiv \mu^2(S_0)\ll 1$. 
We assume (and later verify) that the gravitational and moduli corrections to the vacuum (\ref{ISSvacuum}) around $S_0$ and $T_0$ result in small shifts 
\be
\delta \varphi, \delta \tilde \varphi, \delta \Phi \sim \mu_0^2\,,
\label{shifts}
\ee
for the ISS fields $\varphi,\tilde\varphi$ and $\Phi$. In turn, we will see that the ISS dynamics results in negligible shifts $\delta S,
 \delta T\sim \mu_0^2$ for $S_0$ and $T_0$. In order to be able to estimate the sizes of all terms in $V_{\rm tot}$, we still need
to know the typical scale of the racetrack sector, namely the value of $W_{\rm RT}(S_0,T_0)$.
The relative scale of $W_{\rm RT}$ and $\mu_0$ is fixed by requiring a vanishing cosmological constant.
It is simple to see from eq.s~(\ref{KahlerC}), (\ref{Kahler})--(\ref{Kahler-moduli}) that the only potential term 
in the ISS sector of order $\mu_0^4(S)$ is the $\mu^4$ term itself, so that
\be
W_{\rm RT}(S_0,T_0)\sim \mu_0^2\,.
\ee
We are now ready to perform an expansion of $V_{\rm tot}$ in powers of $\epsilon\equiv \mu_0$.
The leading terms are of order $\mu_0^4$ and define $V_s$:   
\be
V_s = e^{K_{\rm RT}} \Big(K_{\rm RT}^{i\bar \jmath}D_i W_{\rm RT} \overline{D_{j}  W_{\rm RT}} 
-3|W_{\rm RT}|^2+ N_c \mu^4(S) \Big)\,,
\label{V0}
\ee
where $i,j=S,T$. The potential (\ref{V0})  includes all the scalar potential arising from the SUGRA model defined by 
$K_{\rm RT}$ and $W_{\rm RT}$ plus the ISS vacuum energy, independent of the ISS fields.
The strong potential $V_s$ is then responsible for the stabilization of  $S$ and $T$ at VEV's $S_{\mu^4}$ and $T_{\mu^4}$. Notice that the potential $V_s$ is given by the racetrack potential analyzed in Sect.~\ref{racetracksect}, up to the $\mu^4$--term in eq.~\eqref{V0}.  
The presence of the $\mu^4$-piece results in a change of the VEV's $S_0$ and $T_0$ as computed in section~\ref{racetracksect}. An analytical and simple estimate of the VEV's displacements seems  possible only for the RT3 model, in an expansion in derivative with respect to $S$, where  $W_{RT}'''\gg W_{RT}''\gg W_{RT}'$. We demand
\be
\partial_S V_s|_{S_{\mu^4},T_{\mu^4}}=0\qquad , \qquad \partial_T V_s|_{S_{\mu^4},T_{\mu^4}}=0
\ee
and expand up to linear order in $\delta S_0\equiv  S_{\mu^4}-S_0$, $\delta T_0\equiv  T_{\mu^4}-T_0$ around the SUSY vacuum (for $S$ and $T$) $S_0, T_0$, defined, at leading order, by
\be
D_S W_{RT}\simeq W_{RT}^\prime = 0\qquad , \qquad D_TW_{RT}=0 \,.
\ee
In this way, we get 
\bea
\delta S_0&\simeq& -\frac{N_c}{4S_0^2}\left.\frac{(\mu^{4})'}{(W_{RT}'')^2}\right|_{S_0,T_0} \,, \nn\\
\delta T_0&\simeq& \frac{N_c}{4S_0^2}\left.\frac{(\mu^{4})'}{W_{RT}''\partial_T W_{RT}'}\right|_{S_0,T_0}\, .
\eea
Since in the actual models $a_1\simeq a_2\simeq a_3$ and from the cancellation of the cosmological constant $W_{RT}(S_0,T_0)\sim \mu^2(S_0)$ and thus $\eta\simeq a_1$, the above relations yield
\be
\delta S_0\sim -\frac{1}{a_1^3}\ll 1\,, \ \ \ \ \delta T_0\sim \frac{1}{a_1^2}\ll 1\, .
\ee
In the RT2 model, the expansion in derivatives of the dilaton in $W_{RT}$ breaks down.
More precisely, since in the RT2 non--SUSY vacuum $W_{RT}^{\prime\prime}\simeq 0$ (see eq.(\ref{extremaRT2}),
but $W_{RT}^\prime \neq 0$, the above expansion is not consistent with the requirement of the cancellation of the cosmological constant. 
Indeed, a vanishing $V_s$ requires that the terms proportional to $W_{RT}^\prime$ are of the same order as the terms proportional to $W_{RT}$, invalidating the expansion.
A numerical analysis in this case has shown that still $\delta S_0\ll 1$, but $\delta T_0 \sim {\cal O}(1)$.
Thus, the only sizable effect of the $\mu^4$--term in the stabilization of the moduli consists of a shift
of $T$ in the RT2 model .

Let us now turn to $V_w$ and see how the presence of $S$,  as well as gravitational
corrections, modifies the non-SUSY vacuum structure valid at small $\Phi$.
In studying the dynamics of the ISS fields in $V_w$, we can take $S$ and $T$ fixed at their VEV's $S_0$ and $T_0$,
since the dynamics associated to their stabilization is stronger. 
Even with the moduli frozen at their VEV's in $V_w$, a direct analytical study of the potential $V_w(\varphi,\tilde\varphi,\Phi)$ is still quite complicated and probably not very enlightening. It is better to further expand $V_w$ in powers of $\epsilon$. It is not difficult to see that the leading terms
in $V_w$ are of order $\mu_0^6$. As expected, the ${\rm det}\,\Phi$ term in eq.~(\ref{SuperpotC}) is negligible, being of order
$\epsilon_{\rm ISS}^{N_c-2}\mu^{N_c+4}$ and $N_c$ is necessarily greater or equal to 3. 
Further simplifications occur by recalling that $V_s$ is tuned to be vanishing at the minimum.
One finds
\bea
V_w \simeq && \!\!\! e^{K_{\rm RT}} \Bigg[ {\rm Tr} \bigg(|\partial_\Phi W_{\rm ISS}|^2+|\partial_\varphi W_{\rm ISS}+\partial_\varphi K_{\rm ISS} W_{\rm RT}|^2+|\partial_{\tilde\varphi} W_{\rm ISS}+\partial_{\tilde \varphi} K_{\rm ISS}W_{\rm RT}|^2\bigg) \nn \\
&&\qquad+K^{i\bar\jmath}(D_i W_{\rm RT} K_{RT,\bar \jmath}\overline{W_{\rm ISS}}+c.c.) -N_c\mu_0^4 -\mu_0^2W_{\rm RT}{\rm Tr} (\hat \Phi + \hat \Phi^\dagger) \nn \\
&&\qquad -3\,(W_{\rm RT}\overline{W_{\rm ISS}}+c.c.)
+ c\, \mu_0^2 {\rm Tr} |\hat \Phi|^2\Bigg]\,,
\label{ApproxPot}
\eea
where $W_{\rm RT}$, $K_{RT}$ and $D_i W_{RT}$ are simply constants.
Eq.(\ref{ApproxPot}) contains not only all the  $\mu_0^6$ terms of the full potential,
but also further higher order terms. The latter are irrelevant but allow us to write $V_w$ in the compact
form (\ref{ApproxPot}). The last term in eq.~(\ref{ApproxPot}) is the radiatively generated mass term for $\hat\Phi$. The coefficient $c$ is taken as in \cite{Intriligator:2006dd},
 $c=N_c (\ln 4-1)/(8\pi^2)$. Due to the smallness of the gravitational corrections and of the moduli interactions, we expect its actual value to be close to the ISS one. Its precise value is however not important for our considerations.

It is now simple to look for extrema of $V_w$ around the ISS solution (\ref{ISSvacuum}). The gravitational and moduli corrections result only in small shifts in the ISS fields, of the expected order $\mu_0^2$. More precisely, we get (taking all fields as real)
\bea
\delta \hat\Phi_{ij}&=&-\frac{2 N_c^{3/2}}{c \sqrt{3}} \mu_0^2\left[1+\frac{\xi_T T_{\mu^4}}{2}+\xi_S\left( \frac{S_{\mu^4}}{2}+\eta S_{\mu^4}^2\right)\right]\nn\\
\delta Y_{ij} &=&-\frac{\sqrt{N_c}}{\sqrt{3}} \mu_0^2(1+\xi_S\eta S_{\mu^4}^2)\label{shiftsISS}\\
\delta \chi&\sim&\delta\tilde \chi\sim \mu_0^3
\nn
\eea
where we defined $\xi_S$ and $\xi_T$ via $D_SW_{\rm RT}=\xi_S\sqrt{N_c/3}\mu_0^2$ and  $D_TW_{\rm RT}=\xi_T\sqrt{N_c/3}\mu_0^2$. $\xi_S$ and $\xi_T$ are coefficients of ${\cal O}(0.01)$ and ${\cal O}(0.1)$, respectively, in both the RT2 and RT3 models.  Eq.~(\ref{shiftsISS}) is in agreement with what found in \cite{Dudas:2006gr} for a similar context in which the gravitational corrections of the ISS model 
have been studied in presence of the K\"ahler modulus $T$ in type IIB string theory.

The potential $V_w$ slightly affects the $S$ and $T$ stabilization mechanism, resulting
in additional (compared to $\delta S_0,\delta T_0$ from the $\mu^4$-term in $V_s$) small displacements $\delta S=\langle S\rangle - S_{\mu^4}$ , $\delta T=\langle T\rangle - T_{\mu^4}$ in the VEV's for $S$ and $T$ (as well as in their masses). 
It is easy to estimate this displacement by requiring the new minimum to be an extremum of $V_s+V_w$.
At linear order, one has
\be
\frac{\partial V_w}{\partial X^i}\bigg|_{X_0} +  
\frac{\partial^2 V_s}{\partial X_i\partial X_j}\bigg|_{X_0}\delta X_j = 0
\label{generalshifts}
\ee  
where $X_{1,2}=S,T$. From eq.~(\ref{generalshifts}), we estimate that 
\be
\delta S, \delta T \sim \mu_0^2 \ll \delta S_0, \delta T_0 \ll 1\,,
\ee
and thus these additional moduli shifts are totally negligible. 
Shifts in both the moduli and ISS fields much larger than $\mu_0^2$  are expected to arise from quantum corrections to the K\"ahler potentials $K_{RT}$ and $K_{ISS}$, which we are not considering. 
We expect that such corrections, hard to be determined in general, will only result in quantitative
changes but will not alter the above qualitative picture.

\subsubsection{SUSY Vacua}

In addition to the non-SUSY vacuum (\ref{ISSvacuum}), the model presents
several other vacua. It is hard to find all of them, due to the complexity of the potential (\ref{fullPot}). Some of them will most likely appear at Planckian VEV's for the ISS fields and are absent in the global limit. The closest vacua
to the vacuum (\ref{ISSvacuum}) are the usual $N_c$ SUSY vacua (\ref{susyvac1}) in the global limit. 
Let us verify that such vacua are still there once gravitational corrections and the moduli dynamics
are included.  We assume that the racetrack sector and the supergravity corrections shift the VEV's 
\eqref{susyvac1} by at most
\be 
\delta\varphi,\delta\tilde\varphi,\delta\Phi\sim \mu_0^2 \,.
\ee 
This allows us again to expand the full potential in powers of $\epsilon$ around the tree-level SUSY vacua. Writing
\be V_{\rm tot}=\hat V_s+\hat V_w^{(1)}+\hat V_w^{(2)}+\ldots \ee
we get
\be
\hat V_s = e^{K_{\rm RT}} \Big(K_{\rm RT}^{I\bar J}D_I W_{\rm RT} \overline {D_{J} W_{\rm RT} } 
-3|W_{\rm RT}|^2 \Big)\,
\label{V0b}
\ee
\bea V_w^{(1)}&=& e^{K_{\rm RT}} \big[K_{\rm RT}^{i\bar \jmath}D_i W_{\rm RT} \overline{D_{j} W_{\rm ISS}}-3(W_{\rm ISS}\overline W_{\rm RT}+c.c.)\big]\nn\\
V_w^{(2)}&=& e^{K_{\rm RT}} \Bigg[ {\rm Tr} \bigg(|\partial_\Phi W_{\rm ISS}+\partial_\Phi K_{\rm ISS} W_{\rm RT}|^2+|\partial_\varphi W_{\rm ISS}|^2+|\partial_{\tilde\varphi} W_{\rm ISS}|^2\bigg) \nn \\
&&\qquad +K_{\rm RT}^{i\bar\jmath }D_i W_{\rm ISS} \overline{D_{j} W_{\rm ISS}}-3\,|W_{\rm ISS}|^2\Bigg]
\,,\label{potexp2}
\eea
where $V_w^{(1)}$ and $V_w^{(2)}$ contain terms of ${\cal O}(\mu^5)$ and ${\cal O}(\mu^6)$, respectively.
In this expansion we have used that $W_{\rm ISS}\sim \mu_0^3/\epsilon_{\rm ISS}$ around the SUSY vacua. From these expressions we can estimate the shifts of the moduli $\delta S$, $\delta T$ by expanding $\partial_{\chi_i}V_{\rm tot}=0$. At linear order, we get
\be
\frac{\partial V_w^{(1)}}{\partial X^i}\bigg|_{X_0} +  
\frac{\partial^2 V_s}{\partial X_i\partial X_j}\bigg|_{X_0}\delta X_j = 0
\label{generalshifts2}
\ee  
and thus
\be
\delta S, \delta T \sim  \frac{\mu_0}{\epsilon_{\rm ISS}} \ll 1\,.
\ee
Since $\epsilon_{\rm ISS}$ turns out to be not smaller than $10^{-2}$, whereas $\mu_0 \sim 10^{-7}$,
we see that the moduli shifts are again negligible.

The shifts $\delta\varphi,\delta\tilde\varphi,\delta\Phi$ can be determined by demanding the vanishing of the F-terms and expanding them around $\langle\Phi \rangle_{\rm SUSY}$ and $\varphi,\tilde\varphi=0$ up to linear order in $\delta\varphi,\delta\tilde\varphi,\delta\Phi$. This gives, for the example of $\Phi$,
\be
F_\Phi|_{\langle\Phi \rangle_{\rm SUSY}+\delta\Phi}\simeq K_\Phi W_{RT}+\partial_\Phi^2 W_{\rm ISS}\delta \Phi=0\, .
\ee
Since $K_\Phi W_{\rm RT}\sim\mu_0^3/\epsilon_{\rm ISS}^{(N_c-2)/N_c}$ and $\partial_\Phi^2 W_{\rm ISS}\sim\mu_0\epsilon_{\rm ISS}^{(N_c-2)/N_c}$ at $\Phi=\langle\Phi\rangle_{\rm SUSY}$, we get
\be
\delta\Phi\sim-\,\frac{\mu_0^2}{\epsilon_{\rm ISS}^{2(N_c-2)/N_c}}\,.
\ee
Similar results $\sim \mu_0^2$ hold for the fields $\varphi$ and $\tilde\varphi$. 
Interestingly enough, in both the RT2 and RT3 models, these minima are AdS vacua which are SUSY for RT3 and have broken SUSY in the moduli directions for RT2 (however, recall that in RT2 there exists also a fully supersymmetric AdS saddle point which we re-discover here).

\section{Possible Generalizations and Constraints from a String Embedding}\label{general}

We have already mentioned that the ISS K\"ahler potential (\ref{Kahler}) is the most uncertain term in our model. 
Here, we show how to relax the assumption on the modular weights of the ISS sector fields which we implicitly made in writing eq.~\eqref{KahlerC}. 
More generally, in string theory the ISS fields $\Phi$ and $\varphi$, $\tilde\varphi$ will have modular weights $n_\Phi$, $n_\varphi$, $n_{\tilde\varphi}$ different from zero.
Being low--energy composite effective fields, rather than elementary UV fields, their  K\"ahler potential might also have an arbitrary $S$--dependence.
For simplicity, we assume in the following that the possible $S$--dependence is of monomial type in $S+\bar S$, as for the modulus $T$.
The generic K\"ahler potential for the ISS fields reads then (using the $\Z_2$ symmetry $\varphi\leftrightarrow \tilde\varphi$)
\be
K_{\rm ISS}(\varphi,\tilde \varphi,\Phi,S,T) =   \frac{ {\rm Tr} \,\Phi^\dagger \Phi}{(T+\bar T)^{n_\Phi} (S+\bar S)^{m_\Phi}}+
 \frac{ {\rm Tr}(\varphi^\dagger \varphi+\tilde \varphi^\dagger \tilde\varphi)}{(T+\bar T)^{n_\varphi} (S+\bar S)^{m_\varphi} }\,.
\label{KahlerCmod}
\ee
The scalar potential (\ref{potdecomposition}) becomes now much more involved, because of the non--diagonal form of the
K\"ahler metric $g_{I\bar J}$ induced by eq.~(\ref{KahlerCmod}). It is however not difficult to see that all these mixing lead to terms in the potential at least of order $\mu^6$ and thus do not alter the form of $V_s$, the one responsible for the stabilization of the moduli. The only modification induced in $V_s$ from the K\"ahler potential (\ref{KahlerCmod}) comes from the
$g_{\Phi\bar\Phi}$ metric component. At the ${\cal O}(\mu^4)$ level, this amounts in the following replacement in eq.~(\ref{V0}): 
\be
\mu^4\rightarrow (T+\bar T)^{n_\Phi} (S+\bar S)^{m_\Phi} \mu^4 \,.
\label{mu4Mod}
\ee
As we have seen in \ref{nonSUSYvac}, in the RT3 model the $\mu^4$ term provides only a small shift on the VEV's
of $S$ and $T$, whereas in the RT2 model $S$ gets a small shift, but $T$ can get a correction of order one.
Extending the analysis in  \ref{nonSUSYvac} with the replacement (\ref{mu4Mod}), one gets a similar behavior.
The only relevant shift in the moduli occurs for $T$ in the RT2 model. In particular, no qualitative change in the analysis
of section 3 occurs. Once $S$ and $T$ have been stabilized and frozen at their VEV's, in the weaker potential $V_w$,
the K\"ahler potential eq.~\eqref{KahlerCmod} reduces to eq.~(\ref{KahlerC}) with the obvious identifications
\be
\alpha = (2S_0)^{m_\Phi} (2T_0)^{n_\Phi} \,, \ \ \ \ \ \beta= (2S_0)^{m_\varphi} (2T_0)^{n_\varphi} \,.
\ee

Let us now give a closer look to the implicit assumptions we are taking in our construction and on possible constraints coming from an heterotic
string compactification. 
Firstly, we shall discuss shortly the obvious constraints on the rank of the total gauge group coming from the embedding into the $E_8\times E_8^\prime$ or $Spin(32)/\Z_2$ heterotic string theory.
The number of condensing gauge group factors entering in the racetrack sector is clearly bounded by the maximum rank of 16 that we can get in a generic Calabi--Yau compactification of heterotic string theory. Rank one gauge groups like $SU(2)$ are not interesting, because they lead to too low strong coupling scales. Moreover, we assume that the visible sector contains a GUT group such as $SO(10)$ or $SU(5)$, broken to the Standard Model group by some mechanism, such as Wilson lines. Given also the presence of the ISS gauge group $SU(N_c)$, with $N_c\geq 3$, it is clear that no more than 3 gauge groups can realistically be considered for the racetrack sector. Including also the ISS gauge group, the choices for the 4 gauge groups in the RT3 case can range up to $Sp(4)^2\times SU(4)^2$ for an embedding into the $Spin(32)/\Z_2$ heterotic string and $SU(4)\times SU(5)^2$ for an embedding into the $E_8\times E_8^\prime$ theory in the RT2 case. We will study these two cases later on in the numerical examples.

Another assumption we made is the absence of any light exotic matter. This is generally  a rather non-trivial constraint to fulfill in string model--building, but it is clearly a very reasonable one, since
light exotic matter is phenomenologically problematic. 
It essentially implies the existence of a solution for the vanishing of all D and F  term equations
for the exotic matter fields (aside the ISS sector, of course), where the latter are all massive.
We also require that the fermion spectrum is anomaly--free with no need of any Green--Schwarz anomaly cancellation mechanism.  As well known, in D=4 heterotic models, one typically has a single $U(1)$ symmetry giving rise to $U(1)^3$, mixed $U(1)$ non-abelian and $U(1)$ gravitational anomalies. 
These are cancelled by a D=4 generalization \cite{Dine:1987xk} of the usual
D=10 Green-Schwarz (GS) mechanism \cite{Green:1984sg}. In heterotic string models, the GS mechanism is mediated by 
the universal axion field ${\rm Im}\,S$ and it necessarily leads to the appearance of a Fayet--Iliopoulos term 
for the would--be anomalous $U(1)_X$ gauge field. The $D_X$ term either leads to an unacceptable
SUSY breaking at the string scale (with or without breaking the $U(1)_X$ symmetry) or, in most cases, to a spontaneous breaking of the $U(1)_X$ symmetry (again at the string scale). The latter effect leads to a Higgs mechanism in which a combination
of the dilaton multiplet $S$ and a charged (under $U(1)_X$) multiplet are eaten by the $U(1)_X$ vector multiplet to form a massive vector multiplet and leaving at low energy a massless chiral multiplet.  
It is obvious that under such circumstances our analysis does not apply, because there is no decoupling of the $D_X$ and the $F$--term
conditions and, in particular, one has to pay attention to the gauge invariance of the superpotential.
For this reason, we assume that the fermion spectrum of the underlying string model is free of any gauge or mixed gauge
gravitational anomaly, so that no GS mechanism is at work and $S$ is gauge--invariant.
This is not the generic situation, but it is certainly allowed, for instance by simply having a gauge symmetry breaking pattern
with no $U(1)$'s at all (aside the anomaly free hypercharge $U(1)_Y$ factor, of course), a welcome feature given also the global bound on the gauge group rank and 
the pattern of hidden sector we advocate.

The above considerations applies for a general ${\cal N}=1$ D=4 heterotic compactification 
on a Calabi-Yau manifold with no $H$ flux. It is interesting to consider in some more detail what happens in the orbifold limit \cite{Dixon:1985jw}, in which explicit string constructions are available. As far as our analysis is concerned, the most important property arising in an orbifold compactification is the appearance of an exact $SL(2,Z)$ global symmetry
acting on the K\"ahler modulus $T$ as \cite{Ferrara:1989bc}
\be
T\rightarrow \frac{a T -i b}{i c T+d}, \ \ \ \ \ \ a,b,c,d\in Z, \ \ \ ad-bc =1\,,
\label{Tduality}
\ee
which is the low energy manifestation of $T$ duality in the effective field theory.\footnote{Strictly speaking, the 
SUGRA theory is invariant under the whole set of continuous $SL(2,R)$ transformations given by eq.~(\ref{Tduality}).}
Eq.~(\ref{Tduality}) induces a transformation on the K\"ahler potential for the $T$ modulus in eq.~(\ref{Kahler-moduli}), 
with holomorphic parameter $\lambda=3\ln(ic T+d)$, implying a corresponding transformation of the superpotential
\be
W\rightarrow \frac{W}{(ic T+d)^3}\,.
\label{TdualSuperP}
\ee
Charged fields $\Phi_n$ transform under this symmetry in a way which is determined by their modular weights $n_\Phi$ 
appearing in eq.~(\ref{KahlerCmod}):
\be
\Phi\rightarrow (ic T+d)^{-n_\Phi} \Phi\,.
\label{TdualTransf}
\ee

One may notice that the $T$--dependent threshold corrections appearing in the holomorphic gauge kinetic functions (\ref{holoGKF}) violate the $SL(2,Z)$ symmetry (\ref{Tduality}). As is well known, this is not only a problem, but a welcome feature
because such symmetries are typically anomalous and the transformation of the gauge kinetic functions (\ref{holoGKF}) 
is such that to restore the symmetries at the quantum level.
In the most general case, things are more complicated, since these anomalies are cancelled by a combination
of the effect above and of a non-linear transformation of the dilaton $S$, which again mediates a sort of
GS mechanism canceling universal anomalous terms not cancelled by the threshold corrections \cite{Dixon:1990pc,sl2r}.
Similarly to the $U(1)_X$ anomaly discussed above, this leads to a modification of the tree--level dilaton K\"ahler term as follows:
\be
\ln (S+\bar S)\rightarrow \ln \Big[S+\bar S +\delta \ln(T+\bar T) \Big]\,,
\label{SmodT}
\ee 
where $\delta$ is a radiatively generated coefficient, leading to mixing terms between $S$ and $T$. Contrary to the gauge case, no $D_X$ term is generated.  As long as $\delta \ll 1$, eq.~(\ref{SmodT}) does not significantly alter our analysis. Even if $\delta \leq 1$, the modification (\ref{SmodT}) only changes the location of the minima by small amounts without introducing any destabilizing effect. 
The transformation (\ref{TdualSuperP}) poses a non--trivial constraint on the superpotential (\ref{Superpot}). In particular, the symmetry (\ref{Tduality}) is supposed to be spontaneously broken, e.g. by the fields whose VEV give a mass to the charged fields responsible for the effective couplings $A_i$.
Their  modular  transformations (\ref{Tduality}) and that of ${\rm Im}\,S$  should combine to give eq.~(\ref{TdualSuperP}). As we have seen, the modular weights of the ISS fields can be taken essentially arbitrary, without altering the moduli stabilization mechanism, so that they do not pose further constraints.

\section{Numerical Examples} \label{numerics}

Given the complexity of our model, we have found it useful to study directly the full potential (\ref{fullPot}) numerically.
This analysis has allowed us to check the various perturbative expansions performed in subsection 3.2 and, in addition, it provides a more accurate quantitative estimate of various quantities of interest, such as the moduli masses and VEV's, the gravitino mass etc. Given the various uncertainties at hand, mainly in the form of the K\"ahler potential, these estimates should be taken with some care but nevertheless
should give an idea of the relevant ranges of the various quantities.

\subsection{Possible Choices for  $W_{\rm RT}(S,T)$}\label{fullmodel}

We focus our attention onto the two scenarios already discussed in subsection 2.2, involving two or three condensing gauge groups, named
respectively RT2 and RT3. Recall the corresponding superpotentials
\bea
W_{RT2}&  = &  A_2 e^{-a_2 S-\gamma T}-A_1 e^{-a_1 S}\,, \label{RT2}  \\
W_{RT3}& = &  (A_2 e^{-a_2 S}- A_3 e^{-a_3 S})e^{-\gamma T}  - A_1 e^{-a_1 S}\,.
\label{RT23}
\eea
All the main qualitative differences between $W_{RT2}$ and $W_{RT3}$ discussed in subsection 2.2 continue to be valid now, where
in the moduli stabilization one has to consider also the $\mu^4$ term appearing in eq.~(\ref{V0}). The latter term will be responsible
for small shifts in the moduli masses and VEV's, leading to a slight supersymmetry breaking in the $F_S$ and $F_T$ sector in $W_{RT3}$.

\subsection{The Examples}

\TABLE[t]{
\begin{tabular}{| l | l | l |}
\hline&
$\textrm{RT3}$ & $\textrm{RT2}$ \\
&
$Sp(4)^2\times SU(4)^2\times G_{\rm vis}\qquad\qquad$ & $SU(4)\times SU(5)^2\times G_{\rm vis}$ \\
\hline
 $A_1$ &  1/4 & 1/200 \\
 $A_2$ &  3 & 4 \\
$A_3$ &  1/1000 & - - - \\
$N_1$  & 3 & 5 \\
$N_2$  & 3 & 4 \\
$N_3$  & 4 & - - - \\
$\gamma$  & 1 & $0.15$ \\
$N_c$  & 4 & 5 \\
$A_c$ & $3.5\cdot 10^5$ & $1.9\cdot 10^6$ \\\hline
$\langle S\rangle$ & $1.20$ & $1.69$ \\
$\langle T\rangle$ & $1.40$ & $1.57$ \\
$\Lambda_1$ & $2.9\cdot 10^{13}\,{\rm GeV}$ & $3.349\cdot 10^{13}\,{\rm GeV}$ \\
$\Lambda_2$ & $4.2\cdot 10^{13}\,{\rm GeV}$ & $3.353\cdot 10^{13}\,{\rm GeV}$ \\
$\Lambda_3$ & $3.7\cdot 10^{13}\,{\rm GeV}$ & - - - \\
$\Lambda_{\rm ISS}$ & $3.3\cdot 10^{12}\,{\rm GeV}$ & $8.9\cdot 10^{11}\,{\rm GeV}$ \\
$\mu_0$ & $1.2\cdot 10^{11}\,{\rm GeV}$ & $1.0\cdot 10^{11}\,{\rm GeV}$ \\
$\epsilon_{\rm ISS}$ & $0.04$ & $0.12$ \\
$m_s$ & $3.5\cdot 10^6\,{\rm GeV}$ & $2.3\cdot 10^6\,{\rm GeV}$ \\
$m_t$ & $8.6\cdot 10^3\,{\rm GeV}$ & $860\,{\rm GeV}$ \\
$m_\tau$ & $8.3\cdot 10^3\,{\rm GeV}$ & $601\,{\rm GeV}$ \\
$\sqrt{F_S}$ & $6.7\cdot 10^{9}\,{\rm GeV}$ & $1.3\cdot 10^{10}\,{\rm GeV}$ \\
$\sqrt{F_T}$ & $1.0\cdot 10^{11}\,{\rm GeV}$ & $7.2\cdot 10^{10}\,{\rm GeV}$ \\
$\sqrt{F_{\hat\Phi}}$ & $2.4\cdot 10^{11}\,{\rm GeV}$ & $2.3\cdot 10^{11}\,{\rm GeV}$ \\
$m_{3/2}$ & $1.1\cdot 10^3\,{\rm GeV}$ & $0.6\cdot 10^3\,{\rm GeV}$ \\
$\langle V\rangle/3m_{3/2}^2$ & $-0.04$ & $-0.03$ \\
\hline
\end{tabular}
\caption{Input parameters, VEVs, masses and scales for two specific models. $G_{\rm vis}$ denotes
the visible sector gauge group. $\langle S \rangle$, $\langle T \rangle$ and  $\langle V\rangle/3m_{3/2}^2$ are expressed in (reduced) Planck units. See the text for an explanation of all quantities reported.}
\label{tab.1}}

The numerical analysis starts by choosing reasonable sets of the microscopic parameters $A_i$, $a_i=8\pi^2/N_i$, $\gamma$ and $N_c$,
and  searching for extrema in the resulting potential as a function of the real scalar fields. In order to make the numerical study 
of the scalar potential feasible, we make several simplifications in the ISS sector. In particular, we set to zero most of the ISS fields in 
eq.~(\ref{ISSfields}): 
\be
\rho=\tilde \rho = \tilde Z = Z^t = 0 \,, \ \ \ \ \hat \Phi = _{N_c\times N_c}=\hat\Phi \unit_{N_c}\,,
\ee
and take the fields $\chi$, $\tilde \chi$, $Y$ and $\hat\Phi$ real.

In this way, including the two complex fields $S=s+i \sigma$ and $T=t+i \tau$, the scalar potential becomes a function of 8 real scalar fields, whose minimization is a numerically feasible task. 
Once a minimum is found, we calculate the VEV's of all scalar fields and their masses. 
For several choices of the parameters entering in eq.s~(\ref{Kahler})--(\ref{gammaDef}), 
we have numerically verified all the analytical results described in the previous Section. 

For illustrative purposes, we present in Table~\ref{tab.1}  the VEV's for the $S$ and $T$ moduli (at the non--SUSY vacuum (\ref{ISSvacuum})), their physical scalar masses,
the gravitino mass,  as well as the condensing scales of the various gauge groups, $\mu_0$, $\epsilon_{\rm ISS}$, the value of the (approximately cancelled) cosmological constant and the $F$--terms for $S$, $T$ and $\hat \Phi$. All these quantities are given for both the RT2 and RT3 models, in terms of the input parameters $A_{1,2,3}$, $N_{1,2,3,c}$, $\gamma$ and $A_c$. The values $N_1=N_2=3$ for the 
RT3 model appearing in Table ~\ref{tab.1}  are the ``effective'' values defining the parameters
$a_1=a_2 = 8 \pi^2/3$, taking into account that for pure $Sp(2N)$ SYM theories the one--loop $\beta$--function
reads $\beta(g)=-3(N+1) g^3/(16\pi^2)$.
The ``retro--fitting parameter'' $A_c$ is defined in the following way: assuming that the dynamically generated ISS flavour mass in eq.(\ref{flavmasses}) originates from the highest strong coupling scale in the problem ($\Lambda_2$ for both models in Table~\ref{tab.1}), $A_c$ is introduced as
\bea
\mu^2&=&m_f\Lambda_{\rm ISS}\equiv A_c\,\Lambda_{\rm eff.}^3\Lambda_{\rm ISS}=A_c\, \frac{A_2 e^{-\gamma \langle T\rangle}}{N_2}\, e^{-\frac{8\pi^2}{N_\eta}\langle S\rangle}\label{Ac} \\
&&\quad {\rm with}\;N_\eta=\frac{(2N_c-1)N_2}{2N_c+N_2-1}\quad{\rm and}\quad A_c=\left\{\begin{array}{c}1/M^2 \\1/m\end{array}\right.\nn
\eea
for the two cases of eq.~\eqref{flavmasses}, respectively. 

The combined constraints of having i) $S_0,T_0\sim 1$, ii) only moderate tunings in $A_1/A_3$ (RT3) and $A_1/A_2$ (RT2), iii) sufficiently large condensing scales, iv) the rank constraint of the heterotic string (see section~\ref{general}), and v) cancellation of the cosmological constant, lead to a lower bound on the gravitino mass of order 100 GeV, which thus disfavours gauge mediation of SUSY breaking.

It is important to appreciate how constrained are both models.  In the RT3 model, all 5 continuous input parameters are essentially  fixed and we actually think it is remarkable that reasonable values for the $A_i$ and $\gamma$ achieving this goal exist at all. It is also interesting to notice that the light ISS quark mass $m_f$ as computed from $\Lambda_{\rm ISS}$ and $\mu_0$ is in the correct range to be explained by a dynamical mechanism, at least if one uses the first possibility of eq.~\eqref{flavmasses} which gives, using eq.~\eqref{Ac} with $A_c\sim 10^5$, a mass $M\sim 10^{-2}$. A similar analysis applies to the RT2 model, which has the advantage of allowing somewhat
larger values for $S_0$.  However, retro--fitting the light ISS quark mass $m_f$ as computed from $\Lambda_{\rm ISS}$ and $\mu_0$ does not work so well now. 
Even using the first possibility in eq.~(\ref{flavmasses}) , one now gets, from eq.~\eqref{Ac} with $A_c\sim 10^6$, a mass $M\sim 10^{-3}$.
Note, however, that similarly to the discussion of $S$--dependence of the ISS sector K\"ahler potential in Section~\ref{general}, an exponential $S$--dependence of the K\"ahler potential for $\Phi$
\be
{\rm Tr}\,\Phi^\dag\Phi \quad \rightarrow \quad e^{-\delta_\eta(S+\bar S)}{\rm Tr}\,\Phi^\dag\Phi
\ee
would introduce a rescaling of the $\mu^4$-term in the potential similar to eq.~\eqref{mu4Mod}
\be
\mu^4\rightarrow e^{\delta_\eta(S+\bar S)}\mu^4\,  .
\ee
As this would change the relation \eqref{Ac} for the ``retro--fitting parameter'' $A_c$, such an exponential dependence on $S$ might help in improving the retro--fitting. For example, the value $A_c\sim 10^6$ 
is  naturally obtained requiring $e^{-\delta_\eta(S+\bar S)/2}\sim 10^{-6}$. Since $\mu^2=e^{\eta(S+\bar S)/2}\sim 10^{-15}$ and thus $\delta_\eta\ll \eta$, such a correction greatly improves a dynamical explanation of $m_f$, without significantly changing anything else in the analysis.

\section{Life--time of the metastable dS vacuum}
\label{decay}

The study performed so far shows us that the neighbourhood of the non-SUSY vacuum close to $\Phi=0$ and $\chi=\tilde \chi=\mu_0$ consists -- up the possible existence of far-away supergravity induced minima at Planckian VEV's for $\Phi$ and/or $\varphi$ -- of the ISS-style SUSY AdS vacua at $\langle\varphi\rangle=\langle\tilde\varphi\rangle=0$, $\langle\Phi\rangle\gg \mu_0$ and of the usual runaway Minkowski minimum at $S\to\infty$  beyond a barrier separating the non-SUSY minimum from the runaway regime.

From this structure of the minima it is clear that tunneling in the $S$-direction towards infinity is always subdominant compared to tunneling to the nearby ISS--style SUSY AdS vacuum. The former
proceeds from a slightly dS minimum to Minkowski through a high and not too thin wall. The life--time for this process is incredibly long due to the smallness of the vacuum energy $V_{\rm dS}\sim 10^{-120}$ of the de Sitter vacuum describing our universe. The life--time is exponentially shorter than the recurrence time $t_r\sim e^{24\pi^2/V_{\rm dS}}$, but by only a factor which is exponentially smaller than 
$t_r$, resulting in a negligible decay rate \cite{KKLT}.

The decay rate for tunneling to the ISS--style SUSY AdS vacuum  is instead much higher. This is easily seen by estimating the bounce action for the tunneling process. The tunneling path is like the one in \cite{Intriligator:2006dd}, with $S$ and $T$ essentially fixed during the transition.
Denote the real scalar field along the tunneling path by $\phi$. The path is approximately determined by 1) going uphill from the non-SUSY dS vacuum at $\varphi=\tilde\varphi=\mu_0$, $\Phi\approx 0$ towards the barrier top $\varphi=\tilde\varphi=\Phi=0$ and then 2) at $\varphi=\tilde\varphi=0$ down to the SUSY ISS AdS minimum at $\Phi\sim\mu_0/\epsilon_{\rm ISS}^{(N_c-2)/N_c}$. This potential can be approximated to zeroth order by an asymmetric triangular potential, whose bounce action has been calculated exactly  \cite{Duncan:1992ai} (without use of the thin--wall approximation~\cite{Coleman:1977py})
\be 
\tau_{\rm decay}\sim e^B\quad,\quad B=\frac{32\pi^2}{3}\frac{1+c}{(\sqrt{1+c}-1)^4}\frac{\Delta\phi_+^4}{\Delta V_+} \;\;.\label{lifeISS}
\ee
In eq.(\ref{lifeISS})
\be
c=\frac{\Delta V_-}{\Delta V_+} \frac{\Delta \phi_+}{\Delta \phi_-}\simeq (N_c+1)\epsilon_{\rm ISS}^{(N_c-2)/N_c}
\ee
is determined in terms of the two slopes of the triangular potential, $\Delta\phi_+\simeq\mu_0$ and $\Delta\phi_-\simeq\langle\Phi\rangle_{\rm SUSY}\simeq\mu_0/\epsilon_{\rm ISS}^{(N_c-2)/N_c}\gg \mu_0$ denote the distance along the tunneling path between the barrier top and the non-SUSY dS minimum and the barrier top and the SUSY AdS minimum, respectively, while  $\Delta V_\pm\sim \mu_0^4$ denote the corresponding potential differences~\cite{Duncan:1992ai}. 
Notice that $c\sim\epsilon_{\rm ISS}^q$ with $0<q<1$ and thus $\epsilon_{\rm ISS}$ has to be really small in order to get $c\ll 1$, necessary for deriving the result quoted in~\cite{Intriligator:2006dd},
\be
B\sim \frac{1}{c^4}\frac{\Delta\phi_+^4}{\Delta V_+}\sim\frac{\Delta\phi_-^4}{\Delta V_+}\,.
\ee
In our cases, however, $\epsilon_{\rm ISS}=0.01\ldots 0.1$ which implies that $c={\rm O}(1)$ and we have to retain the full expression \eqref{lifeISS}. Plugging in the numerical results for the two example models of Table~\ref{tab.1} we get 
\bea
{\rm RT2:} && \quad B\sim 10^6\,, \nn\\
{\rm RT3:}&& \quad B\sim 4\cdot 10^5\,, \label{lifetimes}
\eea
which, in turn, yields life--times exponentially larger than the present age of the universe.
These decay times are however by far shorter than the decay time in the $S$--direction
and hence are a good estimate for the total life--time.

Finally, we can argue on general grounds that the gravitational corrections  to the decay time  are negligible in our case, where both the potential and the distances in field space are controlled by the parametrically small quantity $\mu_0\ll 1$. 
The relevant point here is that, as long as the thin--wall approximation and thus the notion of a ``bubble'' of the new vacuum in a sea of the old one is not too bad an approximation, the importance of gravity  on the vacuum decay rate is measured by the ratio $\sigma^2/\Delta V$ between the bubble wall tension $\sigma\simeq\int d\phi\sqrt{2 (V(\phi)-V(\phi_-))}$ and  the potential difference between the dS and the AdS minimum $\Delta V=V_+-V_-$~\cite{CDL}. The quality of the thin--wall approximation is specified by $\delta_\phi^{-1}R$, where $\delta_\phi$ is the thickness of the bubble wall, roughly defined as 
the fuzzy region where the bounce solution interpolates from the true to the false vacuum, and $R$ is the radius of the bubble itself. For $\delta_\phi^{-1}R\gg 1$ the thin--wall approximation is a good one, whereas for $\delta_\phi^{-1}R\gsim1$  it is moderately reasonable. 
In both cases, the gravitational corrections are parametrically  controlled by the size of $\sigma^2/\Delta V$ (although the definition of $\sigma$ in the latter case is valid up to factors of order unity). 
This is best seen by noticing that $\sigma^2/\Delta V\propto \rho_{Sch.}/\rho$, where $\rho$
is the actual size of the bubble and $\rho_{Sch.}$ is its Schwarzschild  radius. Gravity is typically
negligible if $\rho\gg \rho_{Sch.}$, whereas it is important for $\rho\lesssim \rho_{Sch.}$.

We studied the tunneling bounce solution for our cases by numerically solving the exact equations of motion without any approximation and found that the condition $\delta_\phi^{-1}R\gsim1$ is valid in explicit examples resembling the RT2 and RT3 model points in parameter space. 

Once checked that the thin--wall approximation is not a too bad an approximation, we can proceed to
estimate $\sigma^2/\Delta V$. Parametrically, we have 
\bea
\sigma\sim\int d\phi\sqrt{2 (V(\phi)-V(\phi_-))}&\sim& \Delta\phi_-\sqrt{\Delta V_+}\sim \mu_0^3\quad , \quad \Delta V_+\sim \mu_0^4\nn\\
&&\quad \Rightarrow\quad\frac{\sigma^2}{\Delta V}\sim \mu_0^2\ll 1\,,
\eea
which implies that gravitational corrections to the tunneling rate are negligible.

\section{Conclusions}\label{concl}

We have shown in this paper how, under certain assumptions, it is possible
to stabilize the dilaton and the universal K\"ahler modulus in a dS/Minkowski vacuum with low energy
Supersymmetry breaking in a class of SUGRA theories which are low energy descriptions of perturbative heterotic vacua on Calabi--Yau three--folds.  We have achieved that by non--perturbative
gauge dynamics, namely multiple gaugino condensates and baryon and meson dynamics
at low energies as described in \cite{Intriligator:2006dd}.  The model is quite constrained
and result in generic quantitative predictions for the moduli and gravitino masses and the pattern of SUSY breaking, summarized in Table 2 for two particular models.

There is a certain amount of fine--tuning
in our construction, typically unavoidable when using racetrack potentials. We have not
attempted to quantify it, but we believe this is sufficiently moderate, as can be seen by 
looking at the input values of our two numerical examples reported in Table~\ref{tab.1}.

The ISS sector provides mainly for an $F$--term uplifting of the vacuum, 
but it is by no means a crucial ingredient. 
Any other sector sufficiently decoupled from the rest of the theory and with 
SUSY broken at some intermediate scale will be fine as 
well \cite{GomezReino:2006dk,Lebedev:2006qq}. 
Similarly, the  K\"ahler stabilization mechanism (or some other mechanism) might be used in place of the racetrack mechanism, or together with it, in more complicated scenarios, although with some
loss of predictivity.

There are several directions in which our study can be generalized and extended.
Considering that the
one--loop holomorphic gauge kinetic functions generally depend not only on $T$, but also
on the non--universal K\"ahler, complex structure and Wilson line moduli, it is conceivable that these moduli can also be stabilized using a racetrack mechanism, as showed to happen 
for $T$. 
One can also relax the assumption of the absence of anomalous $U(1)_X$ factors and generalize
our study, including charged matter as well, along the lines of \cite{FIterm}.
An analysis of the soft terms that can arise from our construction would also be interesting.
Another point that deserves further study is the quantum stability of the moduli stabilization
mechanism.  Provided that the radiative corrections to the K\"ahler potential for the moduli
are small enough, quantum corrections might be under control, since the moduli are essentially stabilized by the racetrack sector, whereas SUSY is mainly broken in the meson direction of the ISS sector.

Probably the most important point to be addressed is to check whether the assumptions 
we have made can actually be realized in a full--fledged string model all together.
The requirement of having several non--abelian hidden gauge groups with rank $> 1$ when
all D--terms (and F--terms for the matter and moduli fields not considered) 
vanish is the one which seems more stringent.
It is clear that attempting to stabilize all moduli by non--perturbative effects  is a much more complicated
task with respect to the use of tree-level flux--induced superpotentials.
But we think it is worth trying, because of the phenomenological appeal of perturbative heterotic string vacua (gauge coupling unification above all) and of their possibility of admitting a perturbative string description in the UV, at least in orbifold limits. 
Besides the obvious technical difficulties one has in studying the stabilization
of all moduli (beyond $S$ and $T$) in this context, the only physical drawback  we see is the possible appearance of additional light moduli, cosmologically disfavoured, with a mass of the order of the gravitino mass or lighter. This should, however, be compared with the typical drawback one has in intersecting brane models of not having gauge coupling unification
and not having (so far) a string description of flux backgrounds at any scale.

\section*{Acknowledgments}

We would like to thank B. Acharya, M. Bertolini, P. Creminelli, G. Dall'Agata, D. Gallego, C. A. Scrucca and M. Trapletti for useful discussions.
This work is partially supported by the European Community's Human
Potential Programme under contracts MRTN-CT-2004-005104,
and by the Italian MIUR under contract PRIN-2005023102.
MS would like to thank the Galileo Galilei Institute for Theoretical
Physics for the hospitality and the INFN for partial support during the
completion of this work.

\end{document}